\documentclass[floats,floatfix,showpacs,amssymb,prd,nofootinbib,superscriptaddress,aps,notitlepage,reprint]{revtex4-1}
\pdfoutput=1
\usepackage{graphicx}
\usepackage[utf8]{inputenc}
\usepackage{graphicx, epsfig, amssymb} %include figure files
\usepackage{amsmath, amsfonts}
\usepackage{bm} %include bold math: \bm{} creates bold letters in math mode
\usepackage[normalem]{ulem} % either use this (simple) or
\usepackage{soul}
\usepackage[linktocpage]{hyperref}
\usepackage[caption=false]{subfig}
\usepackage[usenames]{color}
\def\be{\begin{equation}}
\def\ee{\end{equation}}

\newcommand{\bea}{\begin{eqnarray}}
\newcommand{\eea}{\end{eqnarray}}
\newcommand{\ben}{\begin{enumerate}}
\newcommand{\een}{\end{enumerate}}
\newcommand{\bi}{\begin{itemize}}
\newcommand{\ei}{\end{itemize}}

\def\ga{\mathrel{\raise.3ex\hbox{$>$\kern-.75em\lower1ex\hbox{$\sim$}}}}
	\def\la{\mathrel{\raise.3ex\hbox{$<$\kern-.75em\lower1ex\hbox{$\sim$}}}}

\def\be{\begin{equation}}
\def\ee{\end{equation}}

\def\I_M{{I_{\scriptscriptstyle M\times M}}}

\def\be{\begin{equation}}
\def\ee{\end{equation}}
\def\bea{\begin{eqnarray}}
\def\eea{\end{eqnarray}}
\newcommand{\beq}{\begin{eqnarray}}
\newcommand{\eeq}{\end{eqnarray}}

\newcommand{\beqal}{\begin{eqnarray}\label}
\newcommand{\none}{\end{eqnarray}}
\newcommand{\beqa}{\begin{eqnarray}}
\newcommand{\eeqa}{\end{eqnarray}}

\begin{document}
\title{\large Infrared-finite graviton two-point function in static de Sitter space}

\author{Rafael P. Bernar}\email{rafael.bernar@icen.ufpa.br}
\affiliation{Faculdade de F\'{\i}sica, Universidade
Federal do Par\'a, 66075-110, Bel\'em, Par\'a, Brazil.}

\author{Lu\'is C. B. Crispino}\email{crispino@ufpa.br}
\affiliation{Faculdade de F\'{\i}sica, Universidade
Federal do Par\'a, 66075-110, Bel\'em, Par\'a, Brazil.}

\author{Atsushi Higuchi}\email{atsushi.higuchi@york.ac.uk}
\affiliation{Department of Mathematics, University of York, YO10 5DD, Heslington, York, United Kingdom.}

\begin{abstract}	
We study quantum gravitational perturbations
in the static patch of de Sitter space. In particular, we determine the symplectic inner product of
these perturbations and use it to write down the graviton two-point function in the state analogous to the Bunch-Davies
vacuum in a certain gauge. We find this two-point function to be infrared-finite and time-translation invariant.
\end{abstract}


\date{\today}

\maketitle

%\tableofcontents

%%%%%%%%%%%%%%%%%%%%%%%%%%%%%%%%%%%%%%%%%%%%%%
\section{Introduction}
%%%%%%%%%%%%%%%%%%%%%%%%%%%%%%%%%%%%%%%%%%%%%%

The interest in phenomena in de~Sitter space is increasing recently, especially due to its relevance to the inflationary cosmology~\cite{Kazanas:1980tx,Sato:1980yn,Guth:1980zm,Linde:1981mu,Albrecht:1982wi}, which recently has gained strong support from observation~\cite{BICEP2}. In addition, current observations indicate that our Universe is expanding in an accelerated rate and may approach de Sitter space asymptotically~\cite{riess,perlmutter}. Physics in de Sitter space is also attracting attention because of the dS/CFT correspondence~\cite{strominger}.

The analysis of gravitational perturbations in de~Sitter space is important particularly for the inflationary cosmology, but the infrared (IR) properties of the graviton two-point function in de Sitter space
have remained a source of controversies over the past 30 years.  The main source of these controversies is that
the graviton mode functions natural to the spatially-flat (or Poincar\'e) patch of de~Sitter space behave in a manner similar to those for minimally-coupled massless scalar field~\cite{Ford:1977dj},
which allows no de~Sitter-invariant vacuum state because of IR divergences~\cite{Ford:1977in,Allen:1985ux}.
Ford and Parker found that this similarity leads to IR divergences in the graviton two-point function though they found no IR divergences in the physical quantities they studied~\cite{Ford:1977dj}. (In fact their work deals with a more general Friedmann-Lema\^{\i}tre-Robertson-Walker spacetime.)

However, since linearized gravity has gauge invariance, it is important to determine whether or not these IR divergences are a gauge artifact.
Indeed, the IR divergences and breaking of de~Sitter symmetry they cause in the \emph{free} graviton theory have been shown to be a gauge artifact in the sense that they can be gauged away if we allow
nonlocal gauge transformations~\cite{Higuchi:1986py,Allen:1986dd}.
This point has recently been made clearer by explicit construction of an IR-finite two-point function~\cite{Higuchi:2011vw}.
The authors of Ref.~\cite{Higuchi:2011vw} also pointed out that a local gauge transformation is sufficient to render the two-point function finite in the infrared in a local region of the spacetime. It is also worth noting that the two-point function of the linearized Weyl tensor computed using a de~Sitter non-invariant propagator with an IR cutoff exhibits no IR divergences~\cite{Mora:2012zh} and agrees with the result~\cite{Kouris:2001hz} calculated using the covariant propagator~\cite{Allen:1986tt,Higuchi:2001uv}.   In fact, in a recent gauge-invariant formulation of free gravitons~\cite{Fewster:2012bj} the Weyl-tensor and graviton two-point functions have been shown to be
equivalent in de~Sitter space~\cite{Higuchi:2012vy}.
%This again shows that at the free-theory level there are no IR divergences and that there is a de~Sitter-invariant vacuum %state.
It has also been argued recently~\cite{Morrison:2013rqa} that there is a de~Sitter-invariant Hadamard state for free gravitons defined in a way similar to the scalar case~\cite{Fewster:2012bj}.

Gravitational perturbations in de~Sitter space have been analyzed mainly in the Poincar\'e patch for two reasons.  Firstly, this patch is the most relevant to the inflationary cosmology.  Secondly, the graviton mode functions in this patch are the simplest.  But there have been some works using other patches.
It has been known for some time that in the global patch of de~Sitter space the free graviton field theory has no IR divergences and that there is a de~Sitter-invariant vacuum state~\cite{Higuchi:1991tn} analogous to the Bunch-Davies vacuum~\cite{Bunch:1978yq} for the scalar field theory (see also Ref.~\cite{Chernikov:1968zm}).   As a result there is an IR-finite
graviton two-point function in this patch~\cite{higuchiweeks}.  An IR-finite graviton two-point function has also been
found in the hyperbolic patch of de~Sitter space~\cite{Hawking:2000ee}.  However, there has been little work on quantum gravitational perturbations in the static patch, which is of physical importance because it represents the region causally accessible to an inertial observer.

In this paper we use the formalism developed by Kodama and Ishibashi~\cite{kodamaishibashi} to study quantum gravitational perturbations in the static patch of de~Sitter space.  In particular, we demonstrate that there is an IR-finite graviton two-point function in the Bunch-Davies-like state in this patch.  We emphasize that this two-point function is time-translation invariant unlike that in the global patch~\cite{higuchiweeks}.  Thus, if
linearized gravity is treated as a thermal field theory inside the cosmological horizon~\cite{gibbonshawking}, then one
finds no IR divergences or secular growth of the kind encountered in the Poincar\'e patch. Although it has been shown
that IR divergences are a gauge artifact in the sense mentioned above, it is useful to demonstrate explicitly
that there is an IR-finite
and time-translation invariant graviton two-point function since there are objections to the existence of de~Sitter
invariant Bunch-Davies-like state in de~Sitter space~\cite{Miao:2011ng,Miao:2013isa}.
%(Some authors have also
%claimed that even the retarded graviton Green's function for graviton field in the covariant gauge was
%wrong~\cite{Antoniadis:1986sb,Woodard:2004ut}, but this claim
%has been shown to be unfounded~\cite{Higuchi:2008fu}.)

The rest of this paper is organized as follows.
In Sec.~\ref{sec:perturbationsdesitter}, we give a brief review of the gauge-invariant perturbation formalism, summarizing some properties of the three types of perturbations: scalar, vector and tensor, in the background spacetime, which is de~Sitter space of $n+2$ dimensions. The solutions of the linearized Einstein equations that these three types of perturbations satisfy are presented. (These solutions were obtained previously by Natario and Schiappa~\cite{Natario:2004jd}.) In Sec.~\ref{sec:gravitontwopoint}, we construct the graviton two-point function, starting by the normalization of the modes for each type of perturbations with respect to symplectic inner product.  In particular, we show that the two-point function is IR-finite in a suitably chosen gauge.
In Sec.~\ref{4dimensions} we present the mode functions in the $3+1$ dimensional case explicitly and find a simplified
expression for the graviton two-point function.
 In Sec.~\ref{sec:finalremarks}, we summarize the results found in this paper and discuss their possible implications. In Appendices~\ref{appendix:vector} and \ref{appendix:scalar} we provide some
details of the calculations to normalize the vector- and scalar-type modes, respectively.
In Appendix~\ref{appendix:scalar-2p-function} we compute the two-point function
for the minimally-coupled scalar field, which is discussed for comparison with the graviton case.  In Appendix~\ref{appendix}, we show that the graviton two-point function vanishes identically if one of the points is at the origin.
This result shows clearly that the values of the graviton two-point function themselves have no physical significance.
Throughout this paper we use the metric  signature $-++\cdots +$ and units such that $c=\hbar=1$.

%%%%%%%%%%%%%%%%%%%%%%%%%%%%%%%%%%%%%%%%%%%%%%%%
\section{Gravitational perturbations in the static patch}
%De Sitter spacetime perturbations in static coordinates}%Perturbations in De Sitter spacetime in static coordinates
\label{sec:perturbationsdesitter}
%%%%%%%%%%%%%%%%%%%%%%%%%%%%%%%%%%%%%%%%%%%%%%

%%%%%%%%%%%%%%%%%%%%%%%%%%%%%%%%%%%%%%%%%%%%%%%%%%%%
\subsection{Background Spacetime} \label{subsec:background}
%%%%%%%%%%%%%%%%%%%%%%%%%%%%%%%%%%%%%%%%%%%%%%%%%%%%
In this section we revisit the classical gravitational perturbation studied in~Ref. \cite{Natario:2004jd}.
The background spacetime will be de Sitter in $n+2$ dimensions with $n\geq 2$ and the line element will take the form
\be
ds^2= g_{\mu\nu}dx^{\mu}dx^{\nu}=-(1-\lambda r^2)dt^2+\frac{dr^2}{1-\lambda r^2}+r^2d\sigma_n^2,
\ee
where
\be
d\sigma_n^2 = \gamma_{ij}(x)dx^i dx^j
\ee
is the line element on the $n$-sphere $S^n$. Thus we are working inside the cosmological horizon in the
so-called static coordinate system. We shall put $\lambda=1$ for simplicity. We shall use the notation established in
Refs.~\cite{kodamaishibashi,kis2000}, with the exception of quantities of the background spacetime, for which we use greek indices.   We define the line element of the two-dimensional orbit space by
\be
ds^2_{\textrm{orb}} =g_{ab}dx^a dx^b=-(1-r^2)dt^2+\frac{dr^2}{1-r^2}.
\ee
We denote the covariant derivatives compatible with the full metric represented by the line element $ds^2$,
the two-dimensional metric represented by $ds^2_{\textrm{orb}}$ and the metric on $S^n$ represented with $d\sigma_n^2$
 by $\nabla_\mu$, $D_a$ and $\hat{D}_i$, respectively.  The greek indices are used for spacetime indices running from $0$ to $n+1$, the first latin indices $a,b,c,\ldots$ are for $t$ and $r$
and the $i,j,k,\ldots$ are for $S^n$.  The connection coefficients for $ds^2$, $ds^2_{\textrm{orb}}$ and
$d\sigma_n^2$
 are denoted by $\Gamma^{\alpha}_{\mu\nu}$, $\Gamma^a_{bc}(t,r)$ and $\hat{\Gamma}^i_{jk}(x)$, respectively.

What we will do next in this section is to consider perturbations in the metric, which can be expanded in terms of harmonic tensors of ranks 0, 1, and 2.  These perturbations are called the scalar-, vector- and (rank-2) tensor-type perturbations, respectively.

%%%%%%%%%%%%%%%%%%%%%%%%%%%%%%%%%%%%%%%%%%%%%%%%%%%%
\subsection{Scalar-type perturbations} \label{subsec:scalarpert}
%%%%%%%%%%%%%%%%%%%%%%%%%%%%%%%%%%%%%%%%%%%%%%%%%%%%
The scalar-type perturbations can be expanded in terms of harmonic functions $\mathbb{S}^{(l\sigma)}$ on the $n$-sphere which satisfy
\be
(\hat{\Delta}_n+k_S^2)\mathbb{S}^{(l\sigma)}=0,
\ee
where $\hat{\Delta}_n$ is the Laplace-Beltrami operator on $S^n$. The set of eigenvalues takes the form
\be
k^2_S=l(l+n-1).
\ee
The label $l$ is a non-negative integer and $\sigma$ represents all labels other than $l$. The harmonic modes of the metric perturbation are given by
\bea
h_{ab}^{(S;l \sigma)}&=&f_{ab}^{(l)}\mathbb{S}^{(l\sigma)},\\
h_{ai}^{(S;l \sigma)}&=&r f_a^{(l)}\mathbb{S}^{(l \sigma)}_i,\\
h_{ij}^{(S;l \sigma)}&=&2r^2(\gamma_{ij}H_L^{(l)}\mathbb{S}^{(l\sigma)}+H_T^{(l)}\mathbb{S}_{ij}^{(l \sigma)}),
\eea
where
\bea
\mathbb{S}_{i}^{(l \sigma)}&=&-\frac{1}{k_S}\hat{D}_i\mathbb{S}^{(l \sigma)}, \\
\mathbb{S}_{ij}^{(l \sigma)}&=&\frac{1}{k^2_S}\hat{D}_i \hat{D}_j \mathbb{S}^{(l \sigma)}+\frac{1}{n}\gamma_{ij}\mathbb{S}^{(l \sigma)},
\eea
and the coefficients $f_a^{(l)}$, $f_{ab}^{(l)}$, $H_L^{(l)}$ and $H_T^{(l)}$ are all functions of $t$ and $r$ and are
gauge-dependent quantities. Notice that the tensors $\mathbb{S}_{ij}^{(l\sigma)}$ are chosen to be traceless.

The modes with $l=0,1$ are special cases (and some of the coefficients above are not defined). For $l=0$
%$\mathbb{S}^{(l \sigma)}$ is a constant, and $\mathbb{S}_i^{(l \sigma)}$ and $\mathbb{S}_{ij}^{(l \sigma)}$ are not %defined.
the perturbed spacetime will be spherically symmetric, but the only such solutions to the Einstein equations are Schwarzschild-de~Sitter spacetime by
Birkoff's theorem~\cite{kodamaishibashi}.  Thus, in our case the only perturbation with $l=0$ will be the change in the
background spacetime to introduce a small black hole, which would inevitably be non-perturbative and singular at the
origin.  Hence, we exclude this case.
% This perturbation consists of a time-independent change of the background metric in the form of a shift in the mass %parameter~\cite{kodamaishibashi}.
For $l=1$
%, the variable $H_T^{(l)}$ is not defined. However we can set $H_T^{(1)}=0$ and regard this equation
%as a gauge condition.
one finds that there is no corresponding nonzero gauge-invariant perturbation
as shown in Appendix B of Ref.~\cite{kodamaishibashi}.
Hence we can impose the condition $l\geq 2$.

It can be shown that the perturbations can be related to a master variable $\Phi_S^{(l)}$, which, for the scalar case, obeys the following equation:
\be
\Box \Phi_S^{(l)}-\frac{V_S}{1-r^2}\Phi_S^{(l)}=0, \label{eq_for_Phi_S}
\ee
where the effective potential is given by
\bea
V_S&=&\frac{1-r^2}{4r^2}\left[4l(l+n-1)+n(n-2)\right.\nonumber \\
&&-\left.(n-2)(n-4)r^2\right].
\eea
The $\Box$ is the d'Alembertian operator in the two-dimensional orbit spacetime with line element $ds^2_{\textrm{orb}}$:
\be
\Box = -\frac{1}{1-r^2} \frac{\partial^2\ }{\partial t^2}+ \frac{\partial\ }{\partial r}(1-r^2)\frac{\partial\ }{\partial r}.
\ee
 The procedure to obtain Eq.~(\ref{eq_for_Phi_S}) involves defining gauge-invariant quantities (for modes with $l \geq 2$), which are given in terms of the gauge-dependent quantities by
\bea
F^{(l)} & = & H_L^{(l)}+\frac{1}{n}H_T^{(l)}+\frac{1}{r}D^{a}(rX_a^{(l)}), \label{scalarF}\\
F_{ab}^{(l)} & = & f_{ab}^{(l)}+D_{a}X_{b}^{(l)}+D_{b}X_{a}^{(l)}, \label{scalarFab}
\eea
with
\begin{equation}
X_a^{(l)}=\frac{r}{k_S}\left(f_a^{(l)}+\frac{r}{k_S}D_a H_T^{(l)}\right).
\end{equation}
Then, the functions $F^{(l)}$ and $F_{ab}^{(l)}$, defined by Eqs.~(\ref{scalarF}) and (\ref{scalarFab}), respectively, are given in terms of the master variable $\Phi_S^{(l)}$ as follows:
\begin{eqnarray}
r^{n-2}F^{(l)}&=&\frac{1}{2n}(\Box +2)(r^{n/2}\Phi_S^{(l)}),\label{F_in_Phi_S}\\
r^{n-2}F_{ab}^{(l)}&=&D_a D_b(r^{n/2}\Phi_S^{(l)}) \nonumber \\
&&-\left(\frac{n-1}{n}\Box+\frac{n-2}{n}\right)r^{n/2}\Phi_S^{(l)} g_{ab}.  \label{Fab_in_Phi_S}
\end{eqnarray}
The details for obtaining the master equation in terms of these gauge-invariant quantities are highly involved and can be found in Refs.~\cite{kodamaishibashi,kis2000}.

One can find solutions with Fourier components proportional to $e^{-i\omega t}$ and regular at the origin, which are given by:
\bea
&&\Phi_S^{(\omega l)}(t,r)=A_S^{(\omega l)}e^{-i\omega t} r^{l+n/2}(1-r^2)^{i\omega /2} \nonumber \\
&& \times F\left(\frac{1}{2}(i\omega +l+n-1),\frac{1}{2}(i\omega +l+2);l+\frac{n+1}{2};r^2\right), \nonumber \\
\label{A_S_def}
\eea
where the function $F(\alpha, \beta; \gamma; z)$ is Gauss' hypergeometric function~\cite{abramowitz}.  The normalization constants $A_S^{(\omega l)}$ will be determined later.

%%%%%%%%%%%%%%%%%%%%%%%%%%%%%%%%%%%%%%%%%%%%%%%%%%%%
\subsection{Vector-type perturbations} \label{subsec:vectorpert}
%%%%%%%%%%%%%%%%%%%%%%%%%%%%%%%%%%%%%%%%%%%%%%%%%%%%
The vector-type perturbations are expanded in terms of harmonic vectors $\mathbb{V}^{(l\sigma)}_{i}$, which satisfy
\be
(\hat{\Delta}_n+k^2_V)\mathbb{V}^{(l\sigma)}_{i}=0, \label{vectorsphericaleq1}
\ee
\be
\hat{D}_j \mathbb{V}^{j (l \sigma)}=0. \label{vectorsphericaleq2}
\ee
Here,
\be
k^2_V=l(l+n-1)-1,
\ee
where $l=1,2,...$ and $\sigma$ again represents all labels other than $l$. The metric perturbations of the vector type read
\bea
h_{ab}^{(V;l \sigma)}&=&0,\\
h_{ai}^{(V;l \sigma)}&=&r f_a^{(l)}\mathbb{V}^{(l \sigma)}_{i}, \\
h_{ij}^{(V;l \sigma)}&=&2r^2H_T^{(l)}\mathbb{V}^{(l \sigma)}_{ij},
\eea
with
\be
\mathbb{V}^{(l \sigma)}_{ij}=-\frac{1}{2k_V}(\hat{D}_i\mathbb{V}^{(l \sigma)}_{j}+\hat{D}_{j}\mathbb{V}^{(l \sigma)}_{i}).
\ee
For $l=1$, the tensors $\mathbb{V}_{ij}^{(l \sigma)}$ vanish, rendering the coefficient $F_a^{(l)}$  undefined. In this case one defines a new gauge-invariant quantity and this gives rise to a rotational perturbation, parametrized by a constant, similar to the Myers-Perry solution~\cite{kodamaishibashi, Myers} if the black hole mass is nonzero.  This means that in our case with no black hole, there is no nonzero gauge-invariant vector-type perturbation with $l=1$.

As in the scalar case, we define a gauge-invariant quantity for $l \geq 2$ as follows:
\be
F^{(l)}_a=f^{(l)}_a+\frac{r}{k_V}D_a H_T^{(l)}. \label{gaugeinvariantvector}
\ee
This quantity is related to a master variable $\Phi_V^{(l)}$ by
\be
r^{n-1}F^{(l)a}=\epsilon^{ab}D_b(r^{n/2}\Phi_V^{(l)}),
\ee
where $\epsilon_{ab}$ is the Levi-Civita tensor of the two-dimensional orbit spacetime. The master variable obeys the following wave equation:
\be
\Box \Phi_V^{(l)} - \frac{V_V}{1-r^2}\Phi_V^{(l)}=0, \label{masterV}
\ee
with
\be
V_V=\frac{1-r^2}{r^2}\left[l(l+n-1)+\frac{n(n-2)}{4}(1-r^2)\right].
\ee
The solutions of Eq.~(\ref{masterV}) regular at the origin are
\bea
&&\Phi^{(\omega l)}_V(t,r)=A_V^{(\omega l)}e^{-i\omega t}r^{l+n/2}(1-r^2)^{i\omega/2} \nonumber \\ %\times \nonumber\\
&& \times F\left(\frac{1}{2}(i\omega +l+1),\frac{1}{2}(i\omega +l+n);l+\frac{n+1}{2};r^2\right). \nonumber \\
\label{Phi_V_explicit}
\eea
The normalization constants $A_V^{(\omega l)}$ will be determined later.

%%%%%%%%%%%%%%%%%%%%%%%%%%%%%%%%%%%%%%%%%%%%%%
\subsection{Tensor-type perturbations} \label{subsec:tensorpert}
%%%%%%%%%%%%%%%%%%%%%%%%%%%%%%%%%%%%%%%%%%%%%%
 It is a well-known fact that solutions to Eqs. (\ref{sphericaltensoreq1}), (\ref{sphericaltensoreq2}) and (\ref{sphericaltensoreq3}) do not exist on $S^2$ \cite{regge-wheeler1957, rubinordonez}. A concise proof of this fact can be found in Ref.~\cite{Higuchi:1986wu}. Thus, we do not have tensor-type modes for gravitational perturbations in $3+1$ dimensions.
For $n\geq 3$, the tensor-type perturbations of the metric can be expanded in terms of symmetric harmonic tensors of second rank $\mathbb{T}_{ij}^{(l\sigma)}$. They obey the following equations:
\bea
(\hat{\Delta}_n+k_T^2)\mathbb{T}^{(l\sigma)}_{ij} & = & 0, \label{sphericaltensoreq1}\\
{\mathbb{T}_{i}}^{i(l \sigma)} & = & 0, \label{sphericaltensoreq2}\\
\hat{D}_{j}{\mathbb{T}_{i}}^{j(l \sigma)} & = & 0. \label{sphericaltensoreq3}
\eea
The set of eigenvalues is given by
\be
k_T^2=l(l+n-1)-2.
\ee
The label $l$ is an integer larger than or equal to $2$. The harmonic modes of the metric perturbation are written as
\bea
h_{ab}^{(T;l \sigma)}&=&0, \label{habT1} \\
h_{ai}^{(T;l \sigma)}&=&0, \label{habT2} \\
h_{ij}^{(T;l \sigma)}&=&2r^2H_T^{(l)}\mathbb{T}^{(l \sigma)}_{ij}. \label{habT3}
\eea
The quantity $H_T^{(l)}$ is already gauge-invariant. It is convenient to introduce a new variable $\Phi_T^{(l)}$ by
\be
\Phi_T^{(l)}=r^{n/2}H_T^{(l)}.
\ee
Then the perturbed Einstein equation for $\Phi_T^{(l)}$ reads
\be
\Box \Phi_T^{(l)}-\frac{V_T}{1-r^2}\Phi_T^{(l)}=0, \label{tensoreqbox}
\ee
where the effective potential is
\bea
V_T&=&\frac{1-r^2}{r^2}\left[l(l+n-1)+\frac{n(n-2)}{4} \right. \nonumber \\ %+\right.\nonumber \\
&&-\left. \frac{n(n+2)}{4}r^2\right].
\eea
The solutions of Eq.~(\ref{tensoreqbox}) regular at the origin are given by
\bea
&&\Phi^{(\omega l)}_T(t,r)=A_T^{(\omega l)}e^{-i\omega t}
r^{l+n/2}(1-r^2)^{i\omega /2} \nonumber \\
&&\times F\left(\frac{1}{2}(i\omega+l+n+1),\frac{1}{2}(i\omega +l);l+\frac{n+1}{2};r^2\right), \nonumber \\
\label{Phi_T_explicit}
\eea
where the normalization constants $A_T^{(\omega l)}$ will be determined later.

%%%%%%%%%%%%%%%%%%%%%%%%%%%%%%%%%%%%%%%%%%%%%%%%%%%%
\section{Graviton Two-Point Function}
\label{sec:gravitontwopoint}
%%%%%%%%%%%%%%%%%%%%%%%%%%%%%%%%%%%%%%%%%%%%%%

%%%%%%%%%%%%%%%%%%%%%%%%%%%%%%%%%%%%%%%%%%%%%%%%%%%%
\subsection{Quantization and the Two-Point Function}
%%%%%%%%%%%%%%%%%%%%%%%%%%%%%%%%%%%%%%%%%%%%%%
Let us explain how to construct the physical\footnote{The word ``physical'' is used here in the sense that all gauge degrees of freedom are fixed.} two-point function in a free field theory with gauge invariance such as linearized gravity (see, e.g.\ Refs.~\cite{Fewster:2012bj,Higuchi:2012vy}).  Suppose the theory is described by a Lagrangian density $\mathcal{L}$, where $\mathcal{L}$ is
a local function of $h_{\mu\nu}$ and $\nabla_\lambda h_{\mu\nu}$. (Though we use a symmetric tensor field theory in our
explanation for an obvious reason, the construction works for any other linear field theories.)
If there are only terms quadratic in the derivative $\nabla_\lambda h_{\mu\nu}$ in the Lagrangian, then
the part of the Lagrangian involving $\nabla_\lambda h_{\mu\nu}$ is written as
\be
\mathcal{L}_{\rm der} = \frac{\sqrt{-g}}{2}K^{\lambda\mu\nu\lambda'\mu'\nu'}\nabla_\lambda h_{\mu\nu}
\nabla_{\lambda'}h_{\mu'\nu'},
\ee
where $K^{\lambda\mu\nu\lambda'\mu'\nu'} = K^{\lambda'\mu'\nu'\lambda\mu\nu}= K^{\lambda\nu\mu\lambda'\mu'\nu'}$, then
we define the conjugate momentum current $p^{\lambda\mu\nu}$ by
\be
p^{\lambda\mu\nu} = K^{\lambda\mu\nu\lambda'\mu'\nu'}\nabla_{\lambda'} h_{\mu'\nu'}.
\ee
For any two solutions $h_{\mu\nu}$ and $h_{\mu\nu}^{\prime}$ to the Euler-Lagrange equations
and their conjugate momentum currents
$p^{\lambda\mu\nu}$ and $p^{\prime\lambda\mu\nu}$ we define their symplectic product by
\be
\Omega(h,h^{\prime}) = -\int_\Sigma d\Sigma n_\alpha (h_{\mu\nu}p^{\prime\alpha\mu\nu}
- p^{\alpha\mu\nu}h^{\prime}_{\mu\nu}),
\ee
where $\Sigma$ is a Cauchy surface and $n^\alpha$ is the future-directed unit normal vector to $\Sigma$.
It can readily be shown that $\Omega(h,h^{\prime})$ is independent of the choice of $\Sigma$~\cite{Friedman:1978wla}.

Now, suppose that the symplectic product $\Omega$ is non-degenerate, i.e.\ that there are no solutions $h^{(\textrm{null})}_{\mu\nu}$ satisfying $\Omega(h,h^{(\textrm{null})})=0$ for all solutions $h_{\mu\nu}$.
Suppose further that $h^{(n)}_{\mu\nu}$, where $n$ represents all (continuous and discrete) labels for solutions,
and their complex conjugates $\overline{h^{(n)}_{\mu\nu}}$  form a complete set of solutions such that $\Omega(h^{(n)},h^{(m)}) = 0$ for all $n$ and $m$ --- i.e.\ $\Omega$ is nonzero only between
$h_{\mu\nu}^{(n)}$ and $\overline{h_{\mu\nu}^{(m)}}$ --- and define the inner product of two solutions by
\be
\langle h^{(m)}, h^{(n)}\rangle = i\Omega(\overline{h^{(m)}}, h^{(n)}). \label{matrixM}
\ee
Now, expand the quantum field $\hat{h}_{\mu\nu}(y)$, where $y$ represents all spacetime coordinates, as
\be
\hat{h}_{\mu\nu}(y) = \sum_{n} [a_n h^{(n)}_{\mu\nu}(y) + a_n^\dagger \overline{h^{(n)}_{\mu\nu}}(y)]. \label{expansion}
\ee
Then, the equal-time canonical commutation relations for the operators $\hat{h}_{\mu\nu}(y)$ are equivalent to
\be
[ a_m, a_n^\dagger] = (M^{-1})_{mn},  \label{commutator}
\ee
where $M^{-1}$ is the inverse of the matrix $M^{mn}= \langle h^{(m)},h^{(n)}\rangle$, and
 $[a_m,a_n]=[a_m^\dagger,a_n^\dagger] = 0$.

Unfortunately, linearized gravity cannot be quantized in this manner because the matrix $M$ defined by Eq.~(\ref{matrixM}) is degenerate due to the gauge invariance:  a pure-gauge solution of the form
$h^{(g)}_{\mu\nu} = \nabla_\mu \Lambda_\nu + \nabla_\nu \Lambda_\mu$ has vanishing symplectic product with any solution.  However, if we fix the gauge completely so that the matrix $M$ is non-degenerate when restricted to the
solutions satisfying the gauge conditions, then we can expand the field operator $\hat{h}_{\mu\nu}(y)$
using only the solutions satisfying the gauge conditions in Eq.~(\ref{expansion}) and quantize this field by requiring
the commutation relations given by Eq. (\ref{commutator}).  This procedure is the gauge-fixed version of the gauge-invariant quantization formulated in Ref.~\cite{Fewster:2012bj}.

Note that, if we normalize the solutions in a given gauge by requiring $M^{mn}=\delta^{mn}$ in Eq.~(\ref{matrixM}),
then we have $[a_m,a_n^\dagger]=\delta_{mn}$.  Then, on the vacuum state $|0\rangle$ annihilated by
the operators $a_n$ the two-point function is
\be
\langle 0|\hat{h}_{\mu\nu}(y)\hat{h}_{\mu'\nu'}(y')|0\rangle
= \sum_n h^{(n)}_{\mu\nu}(y)\overline{h^{(n)}_{\mu'\nu'}(y')},  \label{unphysical-2p-function}
\ee
for example.  In the next subsections we normalize the gravitational perturbations found in Sec.~\ref{sec:perturbationsdesitter} so that
we have $M^{mn}=\delta^{mn}$.  This will make the construction of the two-point function straightforward.

%%%%%%%%%%%%%%%%%%%%%%%%%%%%%%%%%%%%%%%%%%%%%%%%%%%%
\subsection{Inner Product} \label{subsec:normalization}
%%%%%%%%%%%%%%%%%%%%%%%%%%%%%%%%%%%%%%%%%%%%%%
With a suitable normalization of the gravitational perturbation $h_{\mu\nu}$ the part of the Lagrangian density involving derivatives of $h_{\mu\nu}$ reads (after some integration by parts)
\bea
\mathcal{L} &=& \sqrt{-g}\Bigg[ \nabla_\mu h^{\mu\lambda}\nabla^\nu h_{\nu\lambda} - \frac{1}{2}\nabla_{\lambda}h_{\mu\nu} \nabla^\lambda h^{\mu\nu} \Bigg.\nonumber \\
&&+ \frac{1}{2}(\nabla^\mu h - 2\nabla^\nu {h^\mu}_\nu)\nabla_\mu h  \nonumber \\
&&+ \Bigg. \ \text{terms involving just} \hspace{.1cm} h_{\mu\nu} \Bigg].
\eea
Hence, the conjugate momentum current is
\begin{eqnarray}
p^{\lambda\mu\nu} & : = & \frac{1}{\sqrt{-g}}\frac{\partial \mathcal{L}}{\partial (\nabla_\lambda h_{\mu\nu})}
\nonumber \\
& = & g^{\lambda \mu} \nabla_\kappa h^{\kappa \nu} + g^{\lambda \nu}\nabla_\kappa h^{\kappa \mu}
- \nabla^\lambda h^{\mu\nu}  \nonumber \\
&& + g^{\mu\nu}(\nabla^\lambda h - \nabla^\kappa {h^\lambda}_\kappa) \nonumber \\
&&- \frac{1}{2}(g^{\lambda \nu}\nabla^\mu h + g^{\lambda \mu} \nabla^\nu h).  \label{conj-momentum}
\end{eqnarray}
Then the inner product (\ref{matrixM}) between two solutions $h^{(m)}_{\mu\nu}$ and $h^{(n)}_{\mu\nu}$ is
\begin{equation}
\langle h^{(m)},h^{(n)}\rangle:= -i\int_{\Sigma} d\Sigma n_\lambda \left(\overline{h^{(m)}_{\mu\nu}}p^{(n)\lambda\mu\nu}-h^{(n)}_{\mu\nu}\overline{p^{(m)\lambda\mu\nu}}\right), \label{inner}
\end{equation}
where the integration is to be carried out on a $t=$\,constant Cauchy surface of the static patch of de~Sitter space.
Next we find the normalization constants such that the inner product (\ref{inner}) is simply $\delta^{mn}$ (which
also involves Dirac's delta function because $\omega$ is a continuous label).  The calculation will closely follow Ref.~\cite{higuchi}.

%%%%%%%%%%%%%%%%%%%%%%%%%%%%%%%%%%%%%%%%%%%%%%%%%%%%
\subsection{Normalization of the tensor-type modes} \label{subsec:normalizationt}
%%%%%%%%%%%%%%%%%%%%%%%%%%%%%%%%%%%%%%%%%%%%%%
For the tensor-type perturbations, which we denote by $h_{\mu\nu}^{(T;\omega l \sigma)}$, we have $p^{(T;\omega l \sigma)\lambda\mu\nu} = -\nabla^\lambda h^{(T;\omega l\sigma)\mu\nu}$  because $h_{\mu\nu}^{(T;\omega l\sigma)}$ given by
Eqs.~(\ref{habT1})-(\ref{habT3}) are transverse  ($\nabla^\mu h_{\mu\nu}=0$) and traceless (${h^\mu}_\mu = 0$).
Noting that $r =1$ is the position of the horizon,  we find the inner product defined by Eq.~(\ref{inner}) to be
\begin{eqnarray}
\langle h^{(T;\omega l \sigma)} , h^{(T;\omega' l' \sigma')} &&\rangle = (\omega+\omega') \lim_{\rho \to 1}\int_{0}^{\rho}dr\frac{r^n}{1-r^2} \nonumber \\
&&\times \int d\Omega_n \overline{h_{ij}^{(T;\omega l \sigma)}}h^{(T;\omega' l' \sigma')ij}, \label{norminnertensor}
\end{eqnarray}
where $d\Omega_n$ integration is over the unit hypersphere $S^n$. Noting that
\begin{equation}
\int d\Omega_n \overline{\mathbb{T}_{ij}^{(l\sigma)}}\mathbb{T}^{(l' \sigma')ij}=\frac{1}{r^4}\delta^{l l'}\delta^{\sigma \sigma'},
\end{equation}
we have
\bea
\langle h^{(T;\omega l \sigma)} , h^{(T;\omega' l' \sigma')} &&\rangle = 4(\omega+\omega')\delta^{l l'}\delta^{\sigma \sigma'} \nonumber\\
&&\times \lim_{\rho \to 1}\int_{0}^{\rho}\frac{dr}{1-r^2}\overline{\Phi^{(\omega l)}_T}\Phi^{(\omega' l)}_T.
\eea
We have to evaluate the following integral:
\begin{equation}
I_{\rho}=\lim_{\rho \rightarrow 1} \int_{0}^{\rho}\frac{dr}{1-r^2}\overline{\Phi^{(\omega l)}_T}\Phi^{(\omega ' l)}_T.
\end{equation}

Using Eq.~(\ref{tensoreqbox}) satisfied by $\Phi^{(\omega l)}_T$ and $\overline{\Phi^{(\omega l)}_T}$, we find
%\begin{equation}
%\left[\frac{\omega ^2}{1-r^2}+\frac{d}{dr}(1-r^2)\frac{d}{dr}\right]
%\Phi^{(\omega l)}_T-\frac{V_T}{f}\Phi^{(\omega l)}_T=0,
%\end{equation}
%we have
\bea
\frac{\omega '^2-\omega ^2}{1-r^2}\overline{\Phi^{(\omega l)}_T}\Phi^{(\omega ' l)}_T&=&\frac{d}{dr}\left[\Phi^{(\omega ' l)}_T(1-r^2)\frac{d}{dr}\overline{\Phi^{(\omega l)}_T}\right. \nonumber \\
&& \left. -\overline{\Phi^{(\omega l)}_T}(1-r^2)\frac{d}{dr}\Phi^{(\omega ' l)}_T\right].
\eea
Integrating the above equation from $0$ to $\rho$ and then taking the limit $\rho \rightarrow 1$, we find
\bea
\lim_{\rho \rightarrow 1}&&\int_{0}^{\rho}\frac{dr}{1-r^2}\overline{\Phi^{(\omega l)}_T}\Phi^{(\omega ' l)}_T=\frac{1}{\omega '^2 - \omega ^2}\lim_{\rho \rightarrow 1}\Bigg[ (1-r^2)  \nonumber \\
&&\times \left. \left(\Phi^{(\omega ' l)}_T\frac{d}{dr}\overline{\Phi^{(\omega l)}_T}-\overline{\Phi^{(\omega l)}_T}\frac{d}{dr}\Phi^{(\omega ' l)}_T\right)\right]_{r=\rho},
\eea
where we have used that $\overline{\Phi^{(\omega l)}_T}(0)=\Phi^{(\omega ' l)}_T(0)=0$.

We can write, for $r \approx 1$~\cite{gradshteyn2007},
%\begin{eqnarray}
%&\Phi^{(\omega l)}_T& \approx \frac{\Gamma(l+\frac{n+1}{2})
%\Gamma(i\omega)A_T^{(\omega l)}}{\Gamma(\frac{1}{2}(l+i\omega))
%\Gamma(\frac{1}{2}(l+i\omega +n+1))}(1-r^2)^{-i\omega /2}+ \nonumber\\
%&+&\frac{\Gamma(l+\frac{n+1}{2})\Gamma(-i\omega)
%A_T^{(\omega l)}}{\Gamma(\frac{1}{2}(l-i\omega))\Gamma(\frac{1}{2}(l-i\omega +n+1))}(1-r^2)^{i\omega /2}.
%\end{eqnarray}
\be
\Phi^{(\omega l)} = A_T^{(\omega l)}
\left[ B_\omega^l (1-r^2)^{-i\omega/2} + \overline{B_\omega^l}(1-r^2)^{i\omega/2}\right],
\ee
where
%Defining
\begin{equation}
B^l_{\omega}=\frac{\Gamma(l+\frac{n+1}{2})\Gamma(i\omega)}{\Gamma(\frac{1}{2}(l+i\omega))\Gamma(\frac{1}{2}(l+i\omega +n+1))}.  \label{tensor-normalization2}
\end{equation}
Then we have
\begin{eqnarray}
I_{\rho}&=&\int_{0}^{\rho}\frac{dr}{1-r^2}\overline{\Phi^{(\omega l)}}\Phi^{(\omega ' l)} \nonumber\\
&=&\frac{i|A_T^{(\omega l)}| ^2}{\omega '+\omega}\left[B^l_{-\omega}B^l_{-\omega '}\exp[\frac{i}{2}(\omega '+\omega)\ln(1-\rho^2)] \right. \nonumber \\
&&-\left. B^l_{\omega}B^l_{\omega '}\exp[-\frac{i}{2}(\omega ' +\omega)\ln(1-\rho^2)]\right] \nonumber\\
&&+\frac{i|A_T^{(\omega l)}| ^2}{\omega '-\omega}\left[B^l_{\omega}B^l_{-\omega '}\exp[\frac{i}{2}(\omega '-\omega)\ln(1-\rho^2)]\right. \nonumber \\
&&-\left. B^l_{-\omega}B^l_{\omega '}\exp[-\frac{i}{2}(\omega '-\omega)\ln(1-\rho^2)] \right],
\end{eqnarray}
noting that $\overline{B^l_{\omega}}=B^l_{-\omega}$.
Dropping the terms rapidly oscillating as functions of $\omega$ and $\omega'$ in the $\rho\to 1$ limit, we find
\begin{equation}
I_{\rho}=\frac{2|A_T^{(\omega l)}| ^2 |B^l_{\omega}| ^2}{\omega '-\omega}\sin{\left[\frac{\omega '-\omega}{2}\ln\left(\frac{1}{1-\rho^2}\right)\right]}.
\end{equation}
Using that
\begin{equation}
\lim_{L \rightarrow \infty}\frac{\sin{Lx}}{x}=\pi\delta(x),
\end{equation}
we have
\begin{equation}
I_1=\lim_{\rho \rightarrow 1}I_{\rho}=2\pi|A_T^{(\omega l)}| ^2|B^l_{\omega}|^2\delta(\omega '-\omega).
\end{equation}
Now, we choose
\bea
|A_T^{(\omega l)}| ^2 & = & \frac{1}{16\pi\omega|B^l_{\omega}|^2},\nonumber \\
& = & \frac{\sinh\pi \omega \left|\Gamma(\frac{1}{2}(l+i\omega))\Gamma(\frac{1}{2}(l+i\omega + n + 1))\right|^2}
{16\pi^2\left|\Gamma(l + \frac{n+1}{2})\right|^2}, \nonumber \\
 \label{tensor-normalization1}
\eea
where we have used
\be
|\Gamma(i\omega)|^2 = \frac{\pi}{\omega\sinh \pi\omega}.
\ee
Then, the inner product between two modes of the tensor type is just
\be
\langle h^{(T;\omega l \sigma)}, h^{(T;\omega' l' \sigma')} \rangle=\delta^{ll'}\delta^{\sigma \sigma'}\delta(\omega-\omega').
\label{delta-condition}
\ee

%%%%%%%%%%%%%%%%%%%%%%%%%%%%%%%%%%%%%%%%%%%%%%%%%%%%
\subsection{Normalization of the vector-type modes} \label{subsec:normalizationv}
%%%%%%%%%%%%%%%%%%%%%%%%%%%%%%%%%%%%%%%%%%%%%%
For the vector-type modes, first let us show that we can choose a gauge such that the components $h_{ij}$ vanish. For a gauge transformation $h_{\mu\nu} \rightarrow h_{\mu\nu}+\nabla_{\mu} \Lambda_{\nu}+\nabla_{\nu} \Lambda_{\mu}$ with
\be
\Lambda_a=0,\hspace{.2cm} \Lambda_{i}=r^2 \phi \mathbb{V}_{i},
\ee
we find
\bea
H_T^{(l)} &\rightarrow& H_T^{(l)}-k_V\phi, \\
f_{a}^{(l)} &\rightarrow& f_{a}^{(l)}+r D_a \phi.
\eea
We can readily see that $F_a^{(l)}$ defined by Eq.~(\ref{gaugeinvariantvector}) is invariant under this gauge transformation. Thus, by letting $\phi=H_T^{(l)}/k_V$ we have $H_T^{(l)} = 0$ and $F_a^{(l)}=f_a^{(l)}$. This choice of gauge leads to
\bea
h_{ai}^{(V;l \sigma)}&=&rF_a^{(l)}\mathbb{V}_i^{(l \sigma)} \nonumber\\
&=&\frac{1}{r^{n-2}}\epsilon_{ab}D^{b}\left[r^{n/2}\Phi^{(l)}_V\right]\mathbb{V}_i^{(l \sigma)},\label{Vhai}\\
h_{ab}^{(V;l \sigma)}&=&0,  \\
h_{ij}^{(V;l \sigma)}&=&0.
\eea
Then, we find the (gauge-invariant) inner product (\ref{inner}) for the vector-type modes as
\bea
\langle h^{(V;\omega l\sigma)},h^{(V;\omega' l' \sigma')}\rangle
%&&\rangle
%&=& i\int d\Omega_n dr r^n \left(\overline{h_{\mu\nu}}p'^{t\mu\nu}-h'_{\mu\nu}\overline{p^{t\mu\nu}}\right) %\nonumber\\
%&=&i\int d\Omega_n dr r^n \left(\overline{h^{\mu\nu}}
%{p'^{t}}_{\mu\nu}-%h'^{\mu\nu}\overline{{p^{t}}_{\mu\nu}}\right)\nonumber\\
& = & 2i\int d\Omega_n dr r^n \nonumber\\
&&\times \left(\overline{h^{(V;\omega l\sigma)bi}}{p^{(V;\omega' l'\sigma')t}}_{bi}\right. \nonumber \\
&& \left. -
h^{(V;\omega' l' \sigma')bi}\overline{{p^{(V;\omega l \sigma)t}}_{bi}}\right),
\label{innerV}
\eea
where $p^{(V;\omega l \sigma)}$ is expressed in terms of $h^{(V;\omega l\sigma)}$ in Eq.~(\ref{pV-expression}).
%\begin{eqnarray}
%{p^{(V)a}}_{bi}
%&=&{\delta^{a}}_{b}g^{\rho\sigma}\nabla_{\rho}h^{(V)}_{\sigma i}-g^{ac}\nabla_{c}h^{(V)}_{bi}\nonumber \\
%& =  & {\delta^{a}}_{b}\left(g^{cd}D_c h^{(V)}_{di}+\frac{n}{r}D^{c}h^{(V)}_{ci}\right)\nonumber \\
%&& -g^{ac}\left(D_c h^{(V)}_{bi}-\frac{D_c r}{r}h^{(V)}_{bi}\right).
%\end{eqnarray}
We substitute Eq.~(\ref{pV-expression})
and use Eq.~(\ref{Vhai}) in Eq.~(\ref{innerV}).  After a cumbersome but straightforward
calculation involving integration by parts with respect to $r$ and the use of the master equation (\ref{masterV}) to eliminate the second-order time derivative of the master variable, we find
\begin{eqnarray}
\langle h^{(V;\omega l \sigma)},h^{(V;\omega' l \sigma')} \rangle
& = & 2(\omega+\omega')\delta^{ll'}\delta^{\sigma\sigma'}(l-1)(l+n) \nonumber \\
&& \times \int \frac{dr}{1-r^2}\overline{\Phi^{(\omega l)}_V}\Phi^{(\omega' l)}_V.
% -\Phi^{(\omega'l)}\partial_t\overline{\Phi^{(\omega l)}}).
\end{eqnarray}
The details of this calculation can be found in Appendix~\ref{appendix:vector}.
We then require that $h^{(V;\omega l\sigma)}_{\mu\nu}$ satisfy the same normalization condition as  $h^{(T;\omega l\sigma)}_{\mu\nu}$,
i.e.\ Eq.~(\ref{delta-condition}), to
determine the normalization constants $A_V^{(\omega l)}$. With the same reasoning as in the tensor case we find
\be
|A_V^{(\omega l)}|^2=\frac{\sinh\pi\omega\left| \Gamma(\frac{1}{2}(i\omega+l+1))\Gamma(\frac{1}{2}(i\omega+ l+ n))\right|^2}{8\pi^2(l-1)(l+n)\left|\Gamma(l+\frac{n+1}{2})\right|^2}. \label{vector-normalization1}
\ee
%with
%\be
%C^{l}_{\omega}=\frac{\Gamma(l+\frac{n+1}{2})\Gamma(i\omega)}
%{\Gamma(\frac{1}{2}(i\omega+l+1))\Gamma(\frac{1}{2}(i\omega+ l+ n))}. \label{vector-normalization2}
%\ee

%%%%%%%%%%%%%%%%%%%%%%%%%%%%%%%%%%%%%%%%%%%%%%%%%%%%
\subsection{Normalization of the scalar-type modes} \label{subsec:normalizations}
%%%%%%%%%%%%%%%%%%%%%%%%%%%%%%%%%%%%%%%%%%%%%%
Now, we shall find the normalization factors $A_S^{(\omega l)}$ for the scalar-type modes.
We first choose a convenient gauge.  Under the gauge transformation with the gauge function $\Lambda_\mu$ given by
\bea
\Lambda_a & = & \psi_a(t,r)\mathbb{S},\\
\Lambda_i & = & \phi(t,r)\mathbb{S}_i,
\eea
one finds that the gauge-dependent functions transform as~\cite{kis2000}
\bea
f_{ab}^{(l)} & \rightarrow & f_{ab}^{(l)} + D_a \psi_b + D_b \psi_a,\\
f_a^{(l)} & \rightarrow & f_a^{(l)} + r D_a \left( \frac{\phi}{r^2}\right) - \frac{k_S}{r}\psi_a,\\
H_T^{(l)} & \rightarrow & H_T^{(l)}-\frac{k_S}{r^2}\phi,\\
H_L^{(l)} & \rightarrow &  H_L^{(l)} + \frac{k_S \phi}{nr^2} + \frac{D^a r}{r}\psi_a.
\eea
Hence by choosing
\bea
\phi & = & \frac{r^2}{k_S}H_T^{(l)},\\
\psi_a & = &  r\left( \frac{1}{k_S}f_a^{(l)} +\frac{r}{k_S^2}D_a H_T^{(l)}\right),
\eea
we can set
the functions $f_a^{(l)}$ and $H_T^{(l)}$ to zero. Then the perturbations will be
\bea
h_{ai}^{(S;l \sigma)}&=0, \\
h_{ab}^{(S;l \sigma)}&=&F_{ab}^{(l)}\mathbb{S}^{(l \sigma)},\\
h_{ij}^{(S;l \sigma)}&=&2r^2\gamma_{ij}F^{(l)}\mathbb{S}^{(l \sigma)},
\eea
where $F^{(l)}$ and $F^{(l)}_{ab}$ are given in terms of the master variable $\Phi_S^{(l)}$ by
Eqs.~(\ref{F_in_Phi_S}) and (\ref{Fab_in_Phi_S}), respectively.

The conserved inner product (\ref{inner}) with the conjugate momentum current defined by Eq. (\ref{conj-momentum}) can be found as
\be
\langle h^{(S;l\sigma)},h'^{(S;l\sigma)}\rangle = -2i\int_{\Sigma}d\Sigma n_a J^a, \label{scalar-inner}
\ee
where the conserved current $J^a$ is given by
\bea
J^a & = & \mathbb{S}^{(l' \sigma')}\overline{\mathbb{S}^{(l \sigma)}}
\left[ \frac{2}{r}D^c r \left(\overline{F^{(l)ab}}F_{bc}^{(l')} - F^{(l')ab}\overline{F^{(l)}_{bc}}\right)\right.  \nonumber \\
&&  - \frac{1}{2}\left(\overline{F^{(l)bc}}D^aF^{(l')}_{bc} - F^{(l')bc}D^a
\overline{F^{l}_{bc}}\right) \nonumber \\
&& \left.+2(2-n)\left(\overline{F^{(l)}}D^a F^{(l')}-F^{(l')}D^a\overline{F^{(l)}}\right)\right]. \label{cons-current-scalar}
\eea

Though it would be possible to express the inner product (\ref{scalar-inner}) in terms of $\Phi_S^{(l)}$ directly in the static coordinate system, it is much easier to do so if we use the Eddington-Finkelstein coordinates and evaluate it
on the future horizon.    Thus, we define the new coordinate
\be
u=t-\frac{1}{2}\log{\frac{1+r}{1-r}}.
\ee
This coordinate ranges over all real values.
The line element of the orbit spacetime becomes
\be
ds^2_{\textrm{orb}}=-(1-r^2)du^2-2dudr.
\ee
We note that a further coordinate transformation, $u=-\log(\rho-\eta)$ and $r=-\rho/\eta$, would result in the
standard metric in the Poincar\'e patch, $ds^2 = \eta^{-2}(-d\eta^2 + d\rho^2 + \rho^2 d\sigma_n^2)$, with
$0 \leq \rho$ and $\eta < 0$.
From this we see that the $r=$\.constant hypersurface with $r > 1$ is almost a
Cauchy surface.  It is not quite a Cauchy surface because the timelike line $\rho=0$ does not intersect it.  However, we expect the data on this hypersuface completely describe the gravitational perturbations because only one point in the future infinity is removed from it. We work under this assumption. We calculate the symplectic inner product only for
the perturbations that tend to zero as $u\to\pm \infty$ so that we can integrate by parts with respect to $u$.  We believe this is
sufficient because perturbations not satisfying this condition can be considered as limiting cases of those satisfying them.

%In the $r\to 1$ limit of this hypersurface is the cosmological horizon,
On the future cosmological horizon we have $ds^2_{\textrm{orb}} = -2dudr$ with $-\infty < u < \infty$.
Hence, if $\Sigma$ is the constant-$r$ hypersurface,
then in the limit $r \to 1$, i.e.\ as it approaches the future cosmological horizon, we have
\be
\lim_{r\to 1}d\Sigma n^{\lambda}
 = d\Omega_n du \left(\frac{\partial\ }{\partial u}\right)^{\lambda}. \label{futurehorizonef}
\ee
Thus, the inner product (\ref{scalar-inner}) can be evaluated on the future cosmological horizon as
\be
\langle h^{(S;l\sigma)},h'^{(S;l\sigma)}\rangle  = -2i\int d\Omega_n du J_u. \label{inner-prod2}
\ee

One can readily see that the first term in the conserved current (\ref{cons-current-scalar}) does not contribute
because, on the horizon, we have $D^a r = -(\partial/\partial u)^a$ and
\be
\left(\partial/\partial u\right)_a D^c r
\left(\overline{F^{(l)ab}}F_{bc}^{(l')} - F^{(l') ab}\overline{F_{bc}^{(l)}}\right) = 0.
\ee
(This equality follows just from the fact that $F^{(l)}_{ab}$ is a symmetric tensor on the two-dimensional orbit spacetime.)
Then, after dropping terms that are total derivatives with respect to $u$, which do not contribute in the integral
(\ref{inner-prod2}), we find that the current  $J_u$ on the horizon can be written as
\bea
J_u & = & \left[ 2\overline{F^{(l)}_{rr}}\partial_u F^{(l')}_{uu} - 2F_{rr}^{(l')}\partial_u \overline{F^{(l)}_{uu}} \right.  \nonumber \\
&&  - 4\overline{F_{rr}^{(l)}}F^{(l')}_{uu} + 4F_{rr}^{(l')}\overline{F^{(l)}_{uu}} \nonumber \\
&& \left. + 2n(n-2)\left(\overline{F^{(l)}}\partial_u F^{(l')}-F^{(l')} \partial_u \overline{F^{(l)}}\right)\right]\nonumber \\
&&\times \overline{\mathbb{S}^{(l\sigma)}}\mathbb{S}^{(l'\sigma')}, \label{J_u}
\end{eqnarray}
where the relation ${F^{(l)a}}_a = -2(n-2)F^{(l)}$, which can readily be verified using Eqs.~(\ref{F_in_Phi_S}) and (\ref{Fab_in_Phi_S}), has been used. On the horizon we find from Eqs.~(\ref{F_in_Phi_S}) and (\ref{Fab_in_Phi_S})
\bea
F^{(l)}_{rr} & = & D_r D_r (r^{n/2}\Phi_S^{(l)}) = \partial_r^2 (r^{n/2}\Phi_{S}^{(l)}),\label{Flab_in_Phi_S1}\\
F^{(l)}_{uu} & = & D_u D_u(r^{n/2}\Phi_S^{(l)}) = (\partial_u^2 - \partial_u)\Phi_{S}^{(l)},\label{Flab_in_Phi_S2}\\
F^{(l)} & = & \frac{r^{2-n}}{2n}(\Box + 2)(r^{n/2}\Phi_S^{(l)}) \nonumber \\
& = & - \frac{1}{2}\left[ (\partial_u + 1)\left( 1+ \frac{2}{n}\partial_r\right)  -\frac{2}{n}\right]\Phi_S^{(l)},
\label{Fl_in_Phi_S}
\eea
where we have used $\Box = 2(\partial_u + 1)\partial_r$ on the horizon.  We substitute these formulae into
Eq.~(\ref{J_u}) and use Eq.~(\ref{eq_for_Phi_S}) satisfied by $\Phi_S^{(l)}$ on the horizon.
%, which becomes
%\be
%\Box \Phi_S^{(l)} = (l+1)(l+n-2)\Phi_S^{(l)}.
%\ee
We then find
\bea
J_u & = & \frac{n-1}{n}l(l-1)(l+n-1)(l+n) \nonumber \\
&& \times (\overline{\Phi^{(l)}_S}\partial_u \Phi_S^{(l')}
- \Phi_S^{(l')}\partial_u\overline{\Phi^{(l)}_S})\overline{\mathbb{S}^{(l\sigma)}}\mathbb{S}^{(l'\sigma')}.
\label{J_u_final}
\eea
Details of this calculation can be found in Appendix~\ref{appendix:scalar}.

The inner product is obtained by substituting Eq.~(\ref{J_u_final}) into Eq.~(\ref{inner-prod2}). This inner product can be rewritten as
%
%So, the inner product between two scalar-type modes will be given by
%\bea
%&&\langle h^{(\omega l \sigma)},h^{(\omega' l' \sigma')}\rangle=i\frac{l(l-1)(n-1)}{n}\times \nonumber\\
%&&\times (l+n-1)(l+n)\delta^{ll'}\delta^{\sigma \sigma'} \times \nonumber \\
%&&\times \int du(\overline{\Phi_S^{(\omega l)}}\partial_u \Phi_S^{(\omega' l)} -\Phi_S^{(\omega' l)}\partial_u %\overline{\Phi_S^{(\omega l)}}),
%\eea
%which can be written as
\bea
\langle h^{(S;\omega l \sigma)},h^{(S;\omega' l' \sigma')}&&\rangle=i\frac{(n-1)l(l-1)(l+n-1)(l+n)}{n} \nonumber\\
&&\times \lim_{r \to 1}\int d\Sigma n^{\lambda}\overline{\mathbb{S}^{(l\sigma)}}\mathbb{S}^{(l'\sigma')} \nonumber\\
&&\times(\overline{\Phi_S^{(\omega l)}}\partial_{\lambda} \Phi_S^{(\omega' l)} -\Phi_S^{(\omega' l)}\partial_{\lambda} \overline{\Phi_S^{(\omega l)}}).
\nonumber \\
\eea
Now, evaluating this on a $t=$\,constant Cauchy surface in the original $tr$ coordinates, we have
\bea
&&\langle h^{(S;\omega l \sigma)},h^{(S;\omega' l' \sigma')}\rangle=i\frac{l(l-1)(l+n-1)(l+n)(n-1)}{n} \nonumber\\
&&\times \delta^{l l'}\delta^{\sigma \sigma'} \int_{0}^{1} \frac{dr}{1-r^2} (\overline{\Phi_S^{(\omega l)}}\partial_t \Phi_S^{(\omega' l)} -\Phi_S^{(\omega' l)}\partial_t \overline{\Phi_S^{(\omega l)}}).
\eea
We then require the same normalization condition as in the tensor case, i.e.\ Eq.~(\ref{delta-condition}). Then the normalization constants $A_S^{(\omega l)}$  defined by Eq.~(\ref{A_S_def}) can be determined as
\bea
&& |A_S^{(\omega l)}|^2 \nonumber \\
&& =
\frac{n\sinh\pi\omega \left|\Gamma(\frac{1}{2}(i\omega+l+2))\Gamma(\frac{1}{2}(i\omega+ l+ n-1))\right|^2}
{2\pi^2(n-1)l(l-1)(l+n-1)(l+n)\left|\Gamma(l+\frac{n+1}{2})\right|^2}.  \nonumber \\
\label{scalar-normalization1}
\eea
%with
%\be
%D^{l}_{\omega}=\frac{\Gamma(l+\frac{n+1}{2})\Gamma(i\omega)}
%{\Gamma(\frac{1}{2}(i\omega+l+2))\Gamma(\frac{1}{2}(i\omega+ l+ n-1))}. \label{scalar-normalization2}
%\ee

%%%%%%%%%%%%%%%%%%%%%%%%%%%%%%%%%%%%%%%%%%%%%%
\subsection{Infrared finite two-point function} \label{subsec:twopointfunction}
%%%%%%%%%%%%%%%%%%%%%%%%%%%%%%%%%%%%%%%%%%%%%%
In this subsection we write down the graviton two-point function in the state analogous to the Bunch-Davies
vacuum in the gauge we have chosen.   Let us first recall the normalized mode functions we obtained.  The tensor-type
modes are
\be
%h_{ab}^{(T:\omega l \sigma)}&=&0,\\
%h_{ai}^{(T;\omega l \sigma)}&=&0, \\
h_{ij}^{(T;\omega l \sigma)} = 2r^{(4-n)/2}\Phi_T^{(\omega l)}\mathbb{T}^{(l \sigma)}_{ij},
\ee
with all other components vanishing, where $\Phi_T^{(\omega l)}$ is given by Eq.~(\ref{Phi_T_explicit}) with
the normalization constants given by Eq.~(\ref{tensor-normalization1}).
%\bea
%&&\Phi^{(\omega l)}_T(r,t)=A_T^{(\omega l)}e^{-i\omega t}
%r^{l+n/2}(1-r^2)^{i\omega /2}\nonumber \\ % \times \nonumber \\
%&&\times F\left(\frac{1}{2}(i\omega+l+n+1),\frac{1}{2}(i\omega +l);l+\frac{n+1}{2};r^2\right). \nonumber \\
%\eea
%The normalization constants $A_T^{(\omega l)}$ are given by Eqs.~(\ref{tensor-normalization1}) and
%(\ref{tensor-normalization2}).
The vector-type modes are given by
\be
h_{ai}^{(V; \omega l \sigma)}
=\frac{1}{r^{n-2}}\epsilon_{ab}D^{b}\left(r^{n/2}\Phi^{(\omega l)}_V\right)\mathbb{V}_i^{(l \sigma)},
\label{V_ai}
\ee
with all other components vanishing. The master variable $\Phi^{(\omega l)}_V$ is given by
Eq.~(\ref{Phi_V_explicit}) with the normalization constant $A_V^{(\omega l)}$ given by Eq.~(\ref{vector-normalization1}).
%\bea
%&&\Phi^{(\omega l)}_V(r,t)=A_V^{(\omega l)}e^{-i\omega t}r^{l+n/2}(1-r^2)^{i\omega/2} \times \nonumber\\
%&& \times F\left(\frac{1}{2}(i\omega +l+1),\frac{1}{2}(i\omega +l+n);l+\frac{n+1}{2};r^2\right), \nonumber \\
%\eea
%where the normalization constants $A_V^{(\omega l)}$ are given by Eqs.~(\ref{vector-normalization1}) and
%(\ref{vector-normalization2}).
Finally, the scalar-type modes are given by
\bea
h_{ai}^{(S;\omega l \sigma)}&=0, \\
h_{ab}^{(S;\omega l \sigma)}&=&r^{2-n}\left\{ D_a D_b(r^{n/2}\Phi_S^{(\omega l)})\right.  \nonumber \\
&&\left. -\left(\frac{n-1}{n}\Box+\frac{n-2}{n}\right)r^{n/2}\Phi_S^{(\omega l)} g_{ab}\right\}\mathbb{S}^{(l \sigma)},
\nonumber \\
\label{S_ab}   \\
h_{ij}^{(S;\omega l \sigma)}&=&\frac{r^{4-n}}{n}\gamma_{ij}(\Box + 2)(r^{n/2}\Phi_S^{(\omega l)})\mathbb{S}^{(l \sigma)},
\eea
where $\Phi_S^{(\omega l)}$ is given by Eq.~(\ref{A_S_def}) with the normalization constants
$A_S^{(\omega l)}$ given by Eq.~(\ref{scalar-normalization1}).
%\bea
%&&\Phi_S^{(\omega l)}(r,t)=A_S^{(\omega l)}e^{-i\omega t}r^{l+n/2}(1-r^2)^{i\omega /2} \times \nonumber \\
%&& \times F\left(\frac{1}{2}(i\omega +l+n-1),\frac{1}{2}(i\omega +l+2);l+\frac{n+1}{2};r^2\right). \nonumber \\
%\eea
%The normalization constants $A_S^{(\omega l)}$ are given by Eqs.~(\ref{scalar-normalization1}) and
%(\ref{scalar-normalization2}).

Let us first examine the low-$\omega$ behavior of the normalized mode functions $h_{\mu\nu}^{(T;\omega l\sigma)}$, $h_{\mu\nu}^{(V;\omega l\sigma)}$  and
$h_{\mu\nu}^{(S;\omega l\sigma)}$, which coincides with the behavior of the master variables
$\Phi_T^{(\omega l)}$, $\Phi_V^{(\omega l)}$ and $\Phi_S^{(\omega l)}$.  We readily find that they all behave like
$\omega^{1/2}$ in the limit $\omega\to 0$ since $l \geq 2$. This is to be contrasted with the behavior of the
normalized minimally-coupled massless-scalar modes, which behave like $\omega^{-1/2}$ for $l=0$, as shown in
Appendix~\ref{appendix:scalar-2p-function}.  It is interesting to note that the normalization constants $A_T^{(\omega l)}$
are the same as those for the minimally-coupled massless-scalar modes for each $l$.  The only difference is that the angular momentum quantum number $l$ is restricted to be greater than or equal to $2$ for the gravitational perturbations whereas in the massless-scalar case it can take the
value $l=0$, which is responsible for the IR divergences as shown in Appendix~\ref{appendix:scalar-2p-function}.

Now, it is well known that the vacuum state with the two-point function~(\ref{unphysical-2p-function}) is unphysical because it will have singularities in the stress-energy tensor on the horizon. This state is analogous to the
Rindler vacuum~\cite{Fulling:1972md} in Minkowski space and the Boulware vacuum~\cite{Boulware:1974dm}
in Schwarzschild spacetime.  A physically acceptable state is the de~Sitter-invariant
Bunch-Davies state~\cite{Bunch:1978yq}, which is
the thermal state with temperature $H/2\pi$~\cite{gibbonshawking}, where $H$ is the Hubble constant.  This state is
analogous to the Hartle-Hawking state~\cite{Hartle:1976tp} in Schwarzschild spacetime. (Strictly speaking, this result
has been shown explicitly only for scalar field, but it is expected that, for example, a general
proof of Kay and Wald~\cite{Kay:1988mu} can be extended to the graviton field with a suitable definition of the
Hadamard state~\cite{Fewster:2012bj}.)

Now, we expand the graviton field $\hat{h}_{\mu\nu}(y)$ as in Eq.~(\ref{expansion}):
\bea
\hat{h}_{\mu\nu}(y) & = & \sum_{P=S,V,T}\sum_{l=2}^\infty \sum_{\sigma}
\int_0^\infty d\omega \left[ a_{ l\sigma}^{(P)}(\omega)
h^{(P;\omega l\sigma)}_{\mu\nu}(y)  \right.  \nonumber \\
&& \ \ \ \ \ \ \  \ \ \ \ \left.
+ a_{l\sigma}^{(P)\dagger}(\omega)\overline{h^{(P;\omega l\sigma)}_{\mu\nu}}(y)\right].
\eea
(There is no tensor-type contribution for $n=2$.)
In the thermal state of temperature $1/2\pi$ --- recall that we have set $H=1$ --- we have
\bea
\langle a^{(P)\dagger}_{l\sigma}(\omega)a^{(P')}_{l'\sigma'}(\omega')\rangle
& = & \frac{1}{e^{2\pi\omega} - 1} \delta^{PP'}\delta^{ll'}\delta^{\sigma\sigma'}\delta(\omega - \omega'),\nonumber \\
\\
\langle a^{(P)}_{l\sigma}(\omega)a^{(P')\dagger}_{l'\sigma'}(\omega')\rangle
& = & \frac{1}{1- e^{-2\pi\omega} }\delta^{PP'}\delta^{ll'}\delta^{\sigma\sigma'}
\delta(\omega - \omega'), \nonumber \\
\eea
with
$\langle a^{(P)}_{l\sigma}(\omega)a^{(P')}_{l'\sigma'}(\omega')\rangle
= \langle a^{(P)\dagger}_{l\sigma}(\omega)a^{(P')\dagger}_{l'\sigma'}(\omega')\rangle = 0$.
Thus, we find the graviton two-point function to be
\bea
\left\langle \hat{h}_{\mu\nu}(y)\hat{h}_{\mu '\nu '}(y') \right\rangle & = &
\sum_{P=S,V,T}\sum_{l=2}^\infty\sum_{\sigma}\int_{0}^{\infty}d\omega \nonumber \\
&& \times \left\{ \frac{1}{e^{2\pi\omega} - 1}\overline{h^{(P;\omega l \sigma)}_{\mu\nu}}(y)
h^{(P;\omega l\sigma)}_{\mu'\nu'}(y')  \right. \nonumber \\
&&\left. + \frac{1}{1- e^{-2\pi \omega}}h^{(P;\omega l \sigma)}_{\mu\nu}(y)
\overline{h^{(P;\omega l\sigma)}_{\mu'\nu'}}(y')\right\}. \nonumber \\
\label{two-point-result}
\eea

As we have seen, all mode functions $h_{\mu\nu}^{(P;\omega l \sigma)}(y)$ tend to zero as $\omega \to 0$ like $\omega^{1/2}$. Hence, the two-point function (\ref{two-point-result}) computed in the Bunch-Davies-like state
is finite in the infrared.  Note that the two-point
function for the minimally-coupled massless scalar field, which takes a similar form, is IR-divergent (even if there were no
thermal factors) because the $l=0$
mode functions behaves like $\omega^{-1/2}$ in the limit $\omega \to 0$ (cf. Appendix~\ref{appendix:scalar-2p-function}).

%%%%%%%%%%%%%%%%%%%%%%%%%%%%%%%%%%%%%%%%%%%%%%
%\subsection{Infrared behavior} \label{subsec:infrared}
%%%%%%%%%%%%%%%%%%%%%%%%%%%%%%%%%%%%%%%%%%%%%%

%%%%%%%%%%%%%%%%%%%%%%%%%%%%%%%%%%%%%%%%%%%%%%%%%%%%%%%%%%%%%%%%%%%%%%%%%%%%
\section{Mode functions and the two-point function in $3+1$ dimensions} \label{4dimensions}
%%%%%%%%%%%%%%%%%%%%%%%%%%%%%%%%%%%%%%%%%%%%%%%%%%%%%%%%%%%%%%%%%%%%%%%%%%%%
In this Section we present some of our results in four dimensions, i.e.\  with harmonic expansion on $S^2$. As we pointed out before, there are no tensor-type modes in $3+1$ dimensions.  The scalar-type modes will be given in terms of the usual scalar spherical harmonics $Y^{(l,m)}(\theta,\phi)$. Note that we have only one additional label other than $l$. Therefore, perturbations of the scalar type in the gauge we have chosen read
\bea
h_{ai}^{(S;\omega l m)}&=0, \\
h_{tt}^{(S;\omega l m)}&=&\frac{Y^{(l,m)}(\theta,\phi)}{2} \nonumber\\
&&\times \left[\partial_t^2+(1-r^2)^2\partial_r ^2\right](r\Phi_S^{(\omega l)}), \label{h_tt_4d}\\
h_{rr}^{(S;\omega l m)}&=&\frac{Y^{(l,m)}(\theta,\phi)}{2} \nonumber\\
&&\times \left[\partial_r^2+\frac{1}{(1-r^2)^2}\partial_t ^2\right](r\Phi_S^{(\omega l)}), \\
h_{tr}^{(S;\omega l m)}&=&Y^{(l,m)}(\theta,\phi) \nonumber\\
&&\times \left[\partial_r \partial_t+\frac{r}{1-r^2}\partial_t \right](r\Phi_S^{(\omega l)}), \\
h_{ij}^{(S;\omega l m)}&=&\frac{r^{2}Y^{(l,m)}(\theta, \phi)}{2}\gamma_{ij}(\Box + 2)(r\Phi_S^{(\omega l)}),
\label{h_ij_4d}
\eea
with $\gamma_{ij}$ given by the usual metric on the $S^2$, described by the line element:
\be
d\sigma_2^2=d\theta^2+\sin^2\theta d\phi^2.
\ee
The function $\Phi_S^{(\omega l)}$ is now:
\bea
&&\Phi_S^{(\omega l)}(t,r)=A_S^{(\omega l)}e^{-i\omega t}r^{l+1}(1-r^2)^{i\omega /2}  \nonumber \\
&& \times F\left(\frac{1}{2}(i\omega +l+1),\frac{1}{2}(i\omega +l+2);l+\frac{3}{2};r^2\right). \nonumber \\
\label{4dPhi_S}
\eea
The normalization constants take a much simpler form:
\be
|A_S^{(\omega l)}|^2=
\frac{\sinh \pi\omega \left|\Gamma(\frac{1}{2}(i\omega+l+2))\Gamma(\frac{1}{2}(i\omega+ l+1)) \right|^2}
{\pi^2(l-1)l(l+1)(l+2) \left|\Gamma(l+\frac{3}{2})\right|^2}.
\label{scalar-normalization1-4dim}
\ee
%with
%\be
%D^{l}_{\omega}=\frac{\Gamma(l+\frac{3}{2})\Gamma(i\omega)}
%{\Gamma(\frac{1}{2}(i\omega+l+2))\Gamma(\frac{1}{2}(i\omega+ l+1))}. \label{scalar-normalization2-4dim}
%\ee

The solutions to Eqs.~(\ref{vectorsphericaleq1}) and (\ref{vectorsphericaleq2}) for the vector harmonics
on the $S^2$ can be written as \cite{regge-wheeler1957, Higuchi:1986wu}:
\be
Y_{i}^{(l,m)}(\theta,\phi)=\frac{\epsilon_{ij}}{\sqrt{l(l+1)}}\partial^{j}Y^{(l,m)}(\theta,\phi),  \label{4d-vec-sph}
\ee
where $\epsilon_{ij}$ is the totally antisymmetric tensor defined by:
\bea
\epsilon_{\theta\theta}&=&\epsilon_{\phi\phi}=0, \\
\epsilon_{\theta\phi}&=&-\epsilon_{\phi\theta}=\sin\theta.
\eea
Then, the vector-type perturbations are given by
\bea
h_{ti}^{(V; \omega l m)}&=&Y^{(l,m)}_{i}(\theta,\phi)(1-r^2)\partial_r\left(r\Phi^{(\omega l)}_V\right), \\
h_{ri}^{(V; \omega l m)}&=&\frac{Y^{(l,m)}_{i}(\theta,\phi)}{1-r^2}\partial_t\left(r\Phi^{(\omega l)}_V\right),
\eea
with all other components vanishing. The master variable $\Phi^{(\omega l)}_V$ is given by
\bea
&&\Phi^{(\omega l)}_V(t,r)=A_V^{(\omega l)}e^{-i\omega t}r^{l+1}(1-r^2)^{i\omega/2}  \nonumber\\
&& \times F\left(\frac{1}{2}(i\omega +l+1),\frac{1}{2}(i\omega +l+2);l+\frac{3}{2};r^2\right), \nonumber \\
\eea
where the normalization constants $A_V^{(\omega l)}$ are now
\be
|A_V^{(\omega l)}|^2=
\frac{\sinh\pi\omega\left|\Gamma(\frac{1}{2}(i\omega+l+1))\Gamma(\frac{1}{2}(i\omega+ l+ 2))\right|^2}{8\pi^2(l-1)(l+2)\left|\Gamma(l+\frac{3}{2})\right|^2}. \label{vector-normalization1-4dim}
\ee
%with
%\be
%C^{l}_{\omega}=D^{l}_{\omega}=\frac{\Gamma(l+\frac{3}{2})\Gamma(i\omega)}
%{\Gamma(\frac{1}{2}(i\omega+l+1))\Gamma(\frac{1}{2}(i\omega+ l+ 2))}. \label{vector-normalization2-4dim}
%\ee
We note that $\Phi_S^{(\omega l)}$ and $\Phi_V^{(\omega l)}$ are essentially the same, with
the precise relation between them being
\be
\Phi_V^{(\omega l)} =\frac{l(l+1)}{8}\Phi_S^{(\omega l)}. \label{Phi_V_Phi_S}
\ee
%in four dimensions, but the normalization constants are, obviously,
%different due to the relation of the master variables with the perturbations.

Next, we simplify our graviton two-point function in $3+1$ dimensions.  Let us first consider the contribution from the
scalar-type modes.  We define the following tensor differential operators motivated by how the mode functions
$h_{\mu\nu}^{(S;\omega l\omega)}$ are given in terms of $\Phi_S^{(\omega l)}$ (see
Eqs.~(\ref{h_tt_4d})-(\ref{h_ij_4d})):
\bea
\mathcal{D}^{(S)}_{tt} &  = & \frac{1}{2}[\partial_t^2 + (1-r^2)^2 \partial_r^2],\label{DS-def1}\\
\mathcal{D}^{(S)}_{rr} & = & \frac{1}{2}\left[ \partial_r^2 + \frac{1}{(1-r^2)^2}\partial_t^2\right],\\
\mathcal{D}^{(S)}_{tr} & = & \partial_r \partial_t + \frac{r}{1-r^2}\partial_t,\\
\mathcal{D}^{(S)}_{ij} & = & \frac{r^2}{2}\gamma_{ij}(\Box + 2), \label{DS-def4}
\eea
with all other components vanishing.  If $\theta'=0$ in $y'=(t',r',\theta',\phi')$, then the contribution to the graviton two-point function (\ref{two-point-result}) with $n=2$ from the scalar-type modes reads
\be
\Delta^{(S)}_{\mu\nu\mu'\nu'}(y,y')
= \mathcal{D}^{(S)}_{\mu\nu}\mathcal{D}^{(S)}_{\mu'\nu'} G(y,y'),
\ee
where
\bea
G(y,y') & = & \sum_{l=2}^\infty  Y^{(l,0)}(\theta'=0,\phi') Y^{(l,0)}(\theta,\phi)
\int_0^\infty d\omega  \nonumber \\
&& \times  \left[ \frac{\overline{r\Phi_S^{(\omega l)}}(t,r)r'\Phi_S^{(\omega l)}(t',r')}
{e^{2\pi \omega} -1} \right. \nonumber \\
&& \left.+  \frac{r\Phi_S^{(\omega l)}(t,r)\overline{r'\Phi_S^{(\omega l)}}(t',r')}{1 -e^{-2\pi\omega}}
\right], \label{Gyy-dash}
\eea
because $Y^{(l,m)}(\theta'=0,\phi') = 0$, unless $m=0$.
We shall find a simplified expression for $G(y,y')$ next.

It is well known~\cite{Bunch:1978yq} that the two-point function for the conformally-coupled massless scalar field (i.e. $M^2 = 2$) is
\be
\Delta^{(c)}(y,y') = \frac{1}{8\pi^2(1-\cos\mu(y,y')+ i\epsilon(t-t'))},
\ee
where $\mu(y,y')$ is the geodesic distance between the two points $y=(t,r,\theta,\phi)$ and $y'=(t',r',\theta',\phi')$ if they are spacelike separated. For timelike separation of the points, $\cos \mu(y,y') = \cosh \mu_T(y,y')$, where $\mu_T(y,y')$ is the timelike
geodesic distance of the two points.  The term $i\epsilon(t-t')$, where $\epsilon$ is an infinitesimal positive number, indicates
how the singularity at $\mu(y,y')=0$ is avoided.  This two-point function can be expressed in the static patch
(by using Appendix~\ref{appendix:scalar-2p-function}) as follows:
\bea
\Delta^{(c)}(y,y')
& = & \sum_{l=0}^\infty\sum_{m=-l}^l \overline{Y^{(l,m)}}(\theta,\phi)Y^{(l,m)}(\theta',\phi')
 \nonumber \\
&& \times \int_0^\infty d\omega |N^{(\omega l)}|^2R_{\omega l}(r)R_{\omega l}(r') \nonumber \\
&& \times \left[ \frac{e^{i\omega (t-t')}}{e^{2\pi\omega} -1}
+ \frac{e^{-i\omega(t-t')}}{1-e^{-2\pi\omega}}\right],
\label{Delta-c}
\eea
where
\bea
|N^{(\omega l)}|^2 & = &
\frac{\sinh\pi\omega}{4\pi^2}  \nonumber \\
&& \times \frac{\left|\Gamma(\frac{1}{2}(i\omega + l + 1))\Gamma(\frac{1}{2}(i\omega + l + 2))
\right|^2}{\left|\Gamma(l+\frac{3}{2})\right|^2},\\
R_{\omega l}(r) & =  & r^l(1-r^2)^{i\omega/2}  \nonumber \\
&& \times F\left( \frac{1}{2}(i\omega + l +1),\frac{1}{2}(i\omega + l + 2); l + \frac{3}{2};r^2\right). \nonumber \\
\eea
We have used the fact that $R_{\omega l}(r)$ and
 $\sum_{m=-l}^l \overline{Y^{(l,m)}}(\theta,\phi)Y^{(l,m)}(\theta',\phi')$ are both real.
Notice that by Eq.~(\ref{scalar-normalization1-4dim}) we have
\be
|A_S^{(\omega l)}|^2 = \frac{4|N^{(\omega l)}|^2}{(l-1)l(l+1)(l+2)}. \label{P_l_Y}
\ee
We multiply Eq.~(\ref{Delta-c}) by $Y^{(l,0)}(\theta',\phi')$ and integrate over $S^2$.
Using Eq.~(\ref{P_l_Y})
we find by the orthonormality of the spherical harmonics
\bea
&& \frac{rr^{\prime}}{2\pi^2(l-1)l(l+1)(l+2)}  \nonumber \\
&& \times \int d\phi'd\theta' \sin\theta' \frac{Y^{(l,0)}(\theta',\phi')}{1-\cos\mu(y,y')+i\epsilon(t-t')} \nonumber \\
&& = \int_0^\infty d\omega \left[ \frac{\overline{\Phi_S^{(\omega l)}}(t,r)\Phi_S^{(\omega l)}(t',r')}
{e^{2\pi \omega} -1}  \right. \nonumber \\
&& \ \ \ \ \left.
+  \frac{\Phi_S^{(\omega l)}(t,r)\overline{\Phi_S^{(\omega l)}}(t',r')}{1 -e^{-2\pi\omega}}
\right]Y^{(l,0)}(\theta,\phi).
%\mathrm{P}_l(\cos\theta).
\eea
Hence, by using the formula
\be
Y^{(l,0)}(\theta,\phi) = \sqrt{\frac{2l+1}{4\pi}}\mathrm{P}_l(\cos\theta),
\ee
we obtain
\bea
&& \frac{(2l+1)rr^{\prime}}{8\pi^3(l-1)l(l+1)(l+2)}  \nonumber \\
&& \times
\int d\phi'd\theta' \sin\theta' \frac{\mathrm{P}_l(\cos\theta')}{1-\cos\mu(y,y') + i\epsilon(t-t')} \nonumber \\
&& = \int_0^\infty d\omega \left[ \frac{\overline{\Phi_S^{(\omega l)}}(t,r)\Phi_S^{(\omega l)}(t',r')}
{e^{2\pi \omega} -1}  \right.\nonumber \\
&& \ \ \ \ \left. +  \frac{\Phi_S^{(\omega l)}(t,r)\overline{\Phi_S^{(\omega l)}}(t',r')}{1 -e^{-2\pi\omega}}
\right]Y^{(l,0)}(\theta'=0,\phi') Y^{(l,0)}(\theta,\phi). \nonumber \\
\eea
By comparing this expression and Eq.~(\ref{Gyy-dash}) we find
\be
G(y,y') = r^{2}r^{\prime 2}\int d\phi' d\theta'\sin\theta' \frac{Q(\theta')}{1-\cos \mu(y,y')+i\epsilon(t-t')},
\label{G-def}
\ee
where
\be
Q(\theta') = \frac{1}{8\pi^3}\sum_{l=2}^\infty
\frac{2l+1}{(l-1)l(l+1)(l+2)} \mathrm{P}_l(\cos\theta').  \label{Q-def}
\ee
It can be shown that this series is convergent for all $\theta'$.
%This series is convergent for all $\theta'$ because $|\mathrm{P}_l(\cos\theta')|$ is bounded by a $l$-independent
%function of $\theta'$:
%\be
%|\mathrm{P}_l(\cos\theta')| \leq \int_{\theta'}^\pi \frac{dt}{\sqrt{2(\cos\theta' - \cos t)}}.
%\ee

Next, let us examine the contribution of the vector-type modes.  If we let $\theta'=0$ again, then it can be shown that
only the modes with $|m|=1$ contribute.  We note first that
\bea
&& \sum_{m=\pm 1} \overline{Y^{(l,m)}}(\theta,\phi)Y^{(l,m)}(\theta',\phi')\nonumber\\
&& = \frac{2l+1}{2\pi l(l+1)} \mathrm{P}_l^1(\cos\theta)\mathrm{P}_l^1(\cos\theta') \cos(\phi-\phi').
\eea
We choose $\phi'=0$.  This means that the $\theta'$-direction and $\phi'$-direction are identified with the
$x'$- and $y'$-directions, respectively, in the cartesian coordinates.  We denote the unit vectors in the
$x'$- and $y'$-directions by $\hat{e}^{(x)}_{i'}$ and $\hat{e}^{(y)}_{i'}$, respectively.
For small $\theta'$ we have~\cite{gradshteyn2007}
\be
P_l^1(\cos\theta') \approx - \frac{l(l+1)}{2}\sin\theta'.
\ee
Then, for $\theta'\to 0$ and $\phi'\to 0$ we find
\be
\epsilon_{i'j'}\partial^{j'}[ P_l^1(\cos\theta')\cos(\phi-\phi')]
\to \frac{l(l+1)}{2}e^{(\phi)}_{i'},
\ee
where
\be
e^{(\phi)}_{i'} = - e^{(x)}_{i'}\sin\phi + e^{(y)}_{i'}\cos\phi. \label{e-phi-def}
\ee
Then, by Eq.~(\ref{4d-vec-sph}) we obtain
 \bea
&& \sum_{m=\pm 1} \overline{Y_i^{(l,m)}}(\theta,\phi)Y_{i'}^{(l,m)}(\theta'=0,\phi'=0)\nonumber\\
&& = \frac{2l+1}{4\pi l(l+1)} \epsilon_{ij}\partial^j \left[\mathrm{P}_l^1(\cos\theta)\hat{e}^{(\phi)}_{i'}\right] \nonumber \\
&&  = \frac{1}{l(l+1)}\epsilon_{ij}\partial^j \frac{\partial\ }{\partial\theta}\left[Y^{(l,0)}(\theta,\phi)Y^{(l,0)}(\theta'=0,\phi'=0)\hat{e}^{(\phi)}_{i'}\right]. \nonumber \\
\label{vec-to-sca}
\eea

We now define the following differential operators motivated by how the vector-type modes are given in terms of
$\Phi_V^{(\omega l)}$:
\bea
\mathcal{D}^{(V)}_t & = & (1-r^2)\partial_r, \label{DV-def1}\\
\mathcal{D}^{(V)}_r & = & \frac{1}{1-r^2}\partial_t. \label{DV-def2}
\eea
Then the contribution of the vector-type modes to our graviton two-point function can be given as
\be
\Delta_{aia'i'}^{(V)}(y,y') = \mathcal{D}^{(V)}_a \mathcal{D}^{(V)}_{a'}F_{ii'}(y,y')
\ee
with all other components vanishing,
where
\bea
F_{ii'}(y,y') & = & \sum_{l=2}^\infty \sum_{m=\pm 1}\overline{Y^{(l,m)}_{i}}(\theta,\phi) Y_{i'}^{(l,m)}(\theta'=0,\phi'=0) \nonumber \\
&& \times \int_0^\infty d\omega \left[ \frac{r\overline{\Phi_V^{(\omega l)}}(t,r)r'\Phi_V^{(\omega l)}(t',r')}
{e^{2\pi \omega} -1} \right. \nonumber \\
&& \left.+  \frac{r\Phi_V^{(\omega l)}(t,r)r'\overline{\Phi_V^{(\omega l)}}(t',r')}{1 -e^{-2\pi\omega}}
\right].
\eea
By substituting Eq.~(\ref{Phi_V_Phi_S}) into this equation, using Eq.~(\ref{vec-to-sca}) and then using the
definition (\ref{Gyy-dash}) of $G(y,y')$ we obtain
\be
F_{ii'}(y,y') = \frac{1}{8}\epsilon_{ij}\partial^j \left[\partial_\theta G(y,y')\hat{e}_{i'}^{(\phi)}\right].
\ee

In summary, if $\theta'=0$ in $y'=(t',r',\theta',\phi')$, then
our graviton two-point function in $3+1$ dimensions is given by
\bea
&& \Delta_{\mu\nu\mu'\nu'}(y,y') \nonumber \\
&&  =  \mathcal{D}^{(S)}_{\mu\nu}\mathcal{D}^{(S)}_{\mu'\nu'}G(y,y') \nonumber \\
&& \ \ \ \ +\frac{1}{2}\delta_{\{\mu}^a \delta_{\nu\}}^i\delta_{\{\mu'}^{a'}\delta_{\nu'\}}^{i'} \mathcal{D}^{(V)}_{a}\mathcal{D}^{(V)}_{a'}\epsilon_{ij}\partial^j \left[\partial_\theta G(y,y')\hat{e}_{i'}^{(\phi)}\right], \nonumber \\
\eea
where $\{...\}$ indicates symmetrization. (This result is independent of the choice of $\phi'$.)  The differential operators $D^{(S)}_{\mu\nu}$ and $D^{(V)}_{a}$
are defined by Eqs.~(\ref{DS-def1})-(\ref{DS-def4}) and Eqs.~(\ref{DV-def1})-(\ref{DV-def2}), respectively,
the function $G(y,y')$ is defined by Eq.~(\ref{G-def}) in terms of the function $Q(\theta')$ defined by
Eq.~(\ref{Q-def}), and the vector $e^{(\phi)}_{i'}$ is defined by Eq.~(\ref{e-phi-def}).

%%%%%%%%%%%%%%%%%%%%%%%%%%%%%%%%%%%%%%%%%%%%%%%%%%%
\section{Concluding Remarks}
\label{sec:finalremarks}
%%%%%%%%%%%%%%%%%%%%%%%%%%%%%%%%%%%%%%%%%%%%%%
In this paper we studied gravitational perturbations in the static patch, i.e.\ inside the cosmological horizon,
of de Sitter space.  In
particular, we used a gauge-invariant formalism to construct the perturbations and found the symplectic inner product among these perturbations and the graviton two-point function with the gauge degrees of freedom fully fixed.  This
two-point function (\ref{two-point-result}) was found to be finite in the infrared because the normalized
perturbations $h^{(P;\omega l \sigma)}_{\mu\nu}$ behave like $\omega^{1/2}$ as $\omega\to 0$.  By construction
this IR-finite two-point function is invariant under a de~Sitter boost which is the time translation with respect
to the timelike Killing vector in the static patch of de~Sitter space.

We note that the IR-divergent two-point function in the Poincar\'e patch grows as a function of time.
The IR-finite two-point function in the global patch~\cite{higuchiweeks} also grows as a function of time if the two points are kept at a fixed physical distance~\footnote{We thank Steve Giddings for pointing this out.}.  In contrast, the IR-finite two-point function obtained in this paper is invariant under
time translation as mentioned above and, hence, does not grow as  a function of time.

There have been many works reporting that de~Sitter invariance is broken due to IR gravitons. For example,
it is claimed in Refs.~\cite{Tsamis:1996qq,Tsamis:1996qm,Tsamis:2008it} that
the Hubble constant would decrease in time because of IR gravitons. (See Refs.~\cite{Garriga:2007zk,Tsamis:2008zz} for a
criticism of these works and the rebuttal.)
%However, this claim has since been
%withdrawn: in the rebuttal to a criticism on their work~\cite{Garriga:2007zk} they simply
%write, ``This work has not been confirmed, nor redone using dimensional regularization''~\cite{Tsamis:2008zz}.  They now
%claim that the Hubble constant would decrease linearly as $H(t) = H(0)(1- Ct)$~\cite{Tsamis:2008it}.
There are also other works finding IR growth of geometrical fluctuations in inflationary
spacetimes~\cite{Giddings:2010nc,Giddings:2011zd,Giddings:2011ze}.  It is also claimed that some coupling constants
change in time in de~Sitter background due to IR divergences of graviton
propagators~\cite{Kitamoto:2012ep,Kitamoto:2012vj,Kitamoto:2013rea,Kitamoto:2014gva}.

On the other hand, there are some works that suggest that even the IR divergences of minimally-coupled massless scalar field have little physical effect in inflationary
cosmology~\cite{Seery:2010kh,Byrnes:2010yc,gerstenlauer:2011ti,Urakawa:2010it,Tanaka:2012wi,Tanaka:2013xe}.
Recently it has been suggested that this conclusion will extend
to linearized gravity~\cite{Tanaka:2014ina}.  One-loop matter effects on the semi-classical Einstein equations have also been studied in detail with the result that the de~Sitter background is stable at least against small metric perturbations~\cite{Frob:2012ui,Frob:2013ht}.

The reported de~Sitter breaking effects in the Poincar\'e patch described above are caused by interactions, but the symmetry breaking mechanism relies heavily on the de~Sitter breaking already present in the propagator
in the Poincar\'e patch due to IR divergences.
Therefore, the IR-finite and time-translation invariant graviton two-point function found in this paper
appears to be in conflict with these
claims of de~Sitter breaking. (We note that the static patch is the part of the Poincar\'e patch that is causally accessible
to a free-falling observer and, hence, is relevant to the inflationary cosmology.)
In this respect we believe that the ``scheme dependence'' in some of the de~Sitter-breaking
results~\cite{Kitamoto:2012zp} should be investigated further.

In resolving  the issue of whether or not there are gauge-invariant de~Sitter breaking effects due to IR gravitons,
it would be useful to develop
perturbation theory for the gravitational field in the covariant point of view.  Some progress has been made in the covariant
analysis of scalar field
theory~\cite{Marolf:2010zp,Marolf:2010nz,Rajaraman:2010xd,Hollands:2010pr,Hollands:2011we,Higuchi:2010xt, Marolf:2011sh}.
It will be interesting to extend these results to perturbation theory for the gravitational field.
As noted in Ref.~\cite{Gibbons:1976pt} the static patch is closely related to the Euclidean quantum field theory, which
in turn is related to the covariant approach to de~Sitter physics.  We believe that our results will be useful
in constructing the interacting field theory of gravity in the static patch, which is both physically relevant and related to
the covariant approach to perturbative quantum gravity in de~Sitter space.
%%%%%%%%%%%%%%%%%%%%%%%%%%%%%%%%%%%%%%%%%%%%%%%%%%%

%%%%%%%%%%%%%%%%%%%%%%%%%%%%%%%%%%%%%%%%%%%%%%
\begin{acknowledgments}
%%%%%%%%%%%%%%%%%%%%%%%%%%%%%%%%%%%%%%%%%%%%%%
We would like to acknowledge Conselho Nacional de Desenvolvimento Cient\'ifico e Tecnol\'ogico (CNPq), Coordena\c{c}\~ao de Aperfei\c{c}oamento de Pessoal de N\'ivel Superior (CAPES) and Marie Curie action NRHEP-295189- FP7-PEOPLE-2011-IRSES for partial financial support.
A. H. and L. C. also acknowledge partial support from the Abdus Salam International Centre for Theoretical Physics through the Visiting Scholar/Consultant Programme and Associates Scheme, respectively. A. H. thanks the Universidade Federal do Pará (UFPA) in Belém for the kind hospitality.
%%%%%%%%%%%%%%%%%%%%%%%%%%%%%%%%%%%%%%%%%%%%%%
\end{acknowledgments}
%%%%%%%%%%%%%%%%%%%%%%%%%%%%%%%%%%%%%%%%%%%%%%

\appendix
%%%%%%%%%%%%%%%%%%%%%%%%%%%%%%%%%%%%%%%%%%%%%%
\section{Calculation of the inner product for the vector-type modes}\label{appendix:vector}
%%%%%%%%%%%%%%%%%%%%%%%%%%%%%%%%%%%%%%%%%%%%%%

As the vector-type perturbations are traceless (${h^\mu}_\mu = 0$), the conjugate momentum current is just:
\bea
p^{(V)\lambda\mu\nu}&=&g^{\lambda\mu}\nabla_{\kappa}h^{(V)\kappa\nu}+g^{\lambda\nu}\nabla_{\kappa}h^{(V)\kappa\mu}\nonumber\\
&&-g^{\mu\nu}\nabla_{\kappa}h^{(V)\lambda\kappa}-\nabla^{\lambda}h^{(V)\mu\nu}.
\eea
The inner product for the vector case can be written as:
\begin{eqnarray}
\langle h^{(V;\omega l \sigma)}&,&h^{(V;\omega' l' \sigma')}\rangle = 2i\int_{\Sigma} d\Omega_n dr r^n \nonumber \\
&&\times \left(\overline{h^{(V;\omega l \sigma)bi}}{p^{(V;\omega ' l' \sigma')t}}_{bi}\right. \nonumber \\
&&\left.-h^{(V;\omega' l' \sigma')bi}\overline{{p^{(V;\omega l \sigma)t}}_{bi}}\right)
\end{eqnarray}
with $p^{(V;\omega l \sigma)}$ given by
%So we have to evaluate the following:
\begin{eqnarray}
{p^{(V;\omega l \sigma)a}}_{bi}&=&{\delta^{a}}_{b}g^{\rho\nu}\nabla_{\rho}h^{(V;\omega l \sigma)}_{\nu i}-g^{ac}\nabla_{c}h^{(V;\omega l \sigma)}_{bi}
\nonumber\\
&=&{\delta^{a}}_{b}\left(g^{cd}D_c h^{(V;\omega l \sigma)}_{di}
+\frac{nD^{c}r}{r}h^{(V;\omega l \sigma)}_{ci}\right)\nonumber \\
&&-g^{ac}\left(D_c h^{(V;\omega l \sigma)}_{bi}-\frac{D_c r}{r}h^{(V;\omega l \sigma)}_{bi}\right).
\label{pV-expression}
\end{eqnarray}
We have
\begin{eqnarray}
&&\overline{h^{(V;\omega l \sigma)bi}}{p^{(V;\omega' l' \sigma')a}}_{bi}=-\frac{\overline{\mathbb{V}^{(l \sigma)i}}\mathbb{V}^{(l' \sigma')}_{i}}{r^{n-2}}D_e\overline{\Omega_V^{(\omega l)}}\epsilon^{be}\nonumber\\
&&\times\left\{{\delta^{a}}_{b}\left[\epsilon_{df}g^{cd}D_c\left(\frac{D^f\Omega_V^{(\omega' l')}}{r^{n-2}}\right)+n\epsilon_{cf}\frac{D^c r D^f \Omega_V^{(\omega' l')}}{r^{n-1}}\right]\right.\nonumber\\
&&\left.-g^{ac}\left[ \epsilon_{bf}D_c\left(\frac{D^f\Omega_V^{(\omega' l')}}{r^{n-2}}\right)-\epsilon_{bf}\frac{D_c r D^f \Omega_V^{(\omega' l')}}{r^{n-1}}\right]\right\},
\end{eqnarray}
where $\Omega_V^{(\omega l)}=r^{n/2}\Phi_V^{(\omega l)}$.  This can be simplified as
\bea
\overline{h^{(V;\omega l \sigma)bi}}&&{p^{(V;\omega' l' \sigma')a}}_{bi}=\frac{\overline{\mathbb{V}^{(l \sigma)i}}\mathbb{V}^{(l' \sigma')}_{i}}{r^{n-2}}D^e\overline{\Omega_V^{(\omega l)}} \nonumber \\
&&\times \left(\frac{D_e D^a\Omega_V^{(\omega' l')}}{r^{n-2}}+2\frac{D_e r D^a\Omega_V^{(\omega' l')}}{r^{n-1}} \right.\nonumber\\
&&\left.-n\frac{D^a rD_e\Omega_V^{(\omega' l')}}
{r^{n-1}}-\frac{D^a r D_e\Omega_V^{(\omega' l')}}{r^{n-1}}\right).
\eea

Now we calculate the integral
\begin{eqnarray}
I(\overline{\Omega_V},\Omega_V')&=&2i\int_{\Sigma} d\Omega_n dr r^n\overline{h^{(V;\omega l \sigma)bi}}{p^{(V;\omega' l' \sigma')t}}_{bi} \\
&=&2i\delta^{ll'}\delta^{\sigma\sigma'}\left[\int_{0}^{1} dr\partial^t\overline{\Omega_V^{(\omega l)}}\right. \nonumber\\
&&\times \left.\left(\frac{\partial_t \partial^t}{r^{n-2}}+\Gamma^{t}_{tr}\frac{\partial^r}{r^{n-2}}\right)\Omega_V^{(\omega' l)}\right.+\int_{0}^{1} dr\partial^r\overline{\Omega_V^{(\omega l)}} \nonumber\\
&&\left.\times\left(\frac{\partial_r \partial^t}{r^{n-2}}+\Gamma^{t}_{rt}\frac{\partial^t}{r^{n-2}}+2\frac{\partial^t}{r^{n-1}}\right)\Omega_V^{(\omega' l)}\right],
\label{I-Omega-Omega}
\end{eqnarray}
where we used the fact that $\int d\Omega_n\overline{\mathbb{V}^{(l \sigma)i}}\mathbb{V}^{(l' \sigma')}_{i}=\frac{1}{r^2}\delta^{ll'}\delta^{\sigma\sigma'}$.

We use the following equation to eliminate the term $\partial_t \partial^t \Omega_V^{(\omega' l)}$
in Eq.~(\ref{I-Omega-Omega}):
% $\Omega_V^{(\omega' l)}$ obeys the following differential equation:
\bea
\frac{\partial_t\partial^t\Omega_V^{(\omega' l)}}{r^{n-2}} & = &
- \partial_r\left(\frac{\partial^r\Omega_V^{(\omega' l)}}{r^{n-2}}\right)
+ 2\frac{\partial^r\Omega_V^{(\omega' l)}}{r^{n-1}}  \nonumber \\
&& + \frac{[l(l+n-1)-n]\Omega_V^{(\omega' l)}}{r^n}.
\eea
We multiply this equation by $\partial^{t}\overline{\Omega_V^{(\omega l)}}$ and integrate with respect to $r$.
We  use integration by parts for the second term, dropping the boundary term because it oscillates rapidly as a function of $\omega$ and $\omega'$
unless $\omega=\omega'$ and hence can be neglected as a distribution of $\omega$
and $\omega'$. %the boundary term will vanish by the same argument used in Appendix \ref{appendix}).
We substitute the resulting expression into Eq.~(\ref{I-Omega-Omega}) and find the inner product as:
\bea
\langle h^{(V;\omega l \sigma)},h^{(V;\omega' l' \sigma')} \rangle
& = & I(\overline{\Omega_V},\Omega'_V)-I(\Omega'_V,\overline{\Omega_V}) \\
%\ee
%that is
%\bea
%\langle h^{(V;\omega l \sigma)}&,&h^{(V;\omega' l' \sigma')} \rangle
& =& 2i\delta^{ll'}\delta^{\sigma\sigma'}(l-1)(l+n)\int_0^1 dr \nonumber \\
&&\times \frac{\Omega_V^{(\omega' l)}\partial^t\overline{\Omega_V^{(\omega l)}}-\overline{\Omega_V^{(\omega l)}}\partial^t\Omega_V^{(\omega' l)}}{r^n}, \nonumber \\
\eea
i.e.
\bea
\langle &h&^{(V;\omega l \sigma)},h^{(V;\omega' l' \sigma')} \rangle=2i\delta^{ll'}\delta^{\sigma\sigma'}(l-1)(l+n) \nonumber \\
&&\times\int_{0}^{1} \frac{dr}{1-r^2}(\overline{\Phi_V^{(\omega l)}}\partial_t\Phi_V^{(\omega' l)}-\Phi_V^{(\omega'l)}\partial_t\overline{\Phi_V^{(\omega l)}}).
\eea
From this equation we find the normalization constants $A_V^{(\omega l)}$ in Eq.~(\ref{vector-normalization1})
in the same way as in the tensor case.
%We already evaluated an integral of this kind (see, for example, Eq. (\ref{norminnertensor})), so we will have the %desired normalizing coefficients.
%%%%%%%%%%%%%%%%%%%%%%%%%%%%%%%%%%%%%%%%%%%%%%
\section{Calculation of the inner product for the scalar-type modes}\label{appendix:scalar}
%%%%%%%%%%%%%%%%%%%%%%%%%%%%%%%%%%%%%%%%%%%%%%
%
%Since gravitons always come from the past horizon and go to the future horizon (in other words there are
%no bound states) we can evaluate the integral in the inner product at the future horizon
%
As we stated in Sec.~\ref{subsec:normalizations}, we evaluate the inner product for the scalar-type modes on the
future horizon.  Let us first derive Eq. (\ref{futurehorizonef}) on the future horizon.
A future-pointing vector orthogonal to an $r=$\.constant hypersurface, which is spacelike if $r>1$, is $-\nabla_a r$.
Then, the unit future pointing normal vector is
\beqa
n^a & = & (r^2 -1)^{-1/2} \nabla^a r \nonumber \\
& = & (r^2-1)^{-1/2}\left(\frac{\partial\ }{\partial u}\right)^a +
(r^2-1)^{\frac{1}{2}}\left(\frac{\partial\ }{\partial r}\right)^a.
\eeqa
Now,  the surface element of this hypersurface is
\beq
d\Sigma = d\Omega_n du (r^2-1)^{1/2}.
\eeq
Hence
\beq
d\Sigma n^a = d\Omega_n du \left[ \left(\frac{\partial\ }{\partial u}\right)^a
+ (r^2-1)\left(\frac{\partial\ }{\partial r}\right)^a\right].
\eeq
Clearly, in the limit $r\to 1$ we have
\beq
\lim_{r\to 1}d\Sigma n^a
 = d\Omega_n du \left(\frac{\partial\ }{\partial u}\right)^a,
\eeq
which is Eq.~(\ref{futurehorizonef}).

Now, to express the conserved current in terms of the master variable $\Phi_S^{(l)}$, we first simplify Eq.~(\ref{Fl_in_Phi_S}), which expresses $F^{(l)}$ in terms of $\Phi_S^{(l)}$, using the field equation
(\ref{eq_for_Phi_S}), which reads on the horizon
\be
\Box \Phi_S^{(l)} = 2(\partial_u + 1)\partial_r\Phi_S^{(l)} =  A_{n,l}\Phi_S^{(l)},  \label{horizon-field-eq}
\ee
with
\be
A_{n,l}=(l+1)(l+n-2),
\ee
as
\be
F^{(l)}=-\frac{1}{2}\left(\partial_u+1-\frac{A_{n,l}}{n}-\frac{2}{n}\right)\Phi_S^{(l)}.
\ee
Then we find
\bea
\overline{F^{(l)}}\partial_u F^{(l')} &=& \frac{1}{4}\left(\partial_u +1-\frac{2}{n}-\frac{A_{n,l}}{n}\right)\overline{\Phi_S^{(l)}} \nonumber\\
&&\times \partial_u\left(\partial_u +1-\frac{2}{n}-\frac{A_{n,l}}{n}\right)\Phi_S^{(l')} \nonumber\\
&\approx&-\frac{\partial_u \Phi_S^{(l')}}{4} \nonumber\\
&&\times \left[\partial^2 _u-\left(1-\frac{2}{n}-\frac{A_{n,l}}{n}\right)^2\right]\overline{\Phi_S^{(l)}}.
\eea
Here we indicated the equivalence up to a total derivative with respect to $u$ by $\approx$ because we will
integrate this quantity over $u$ to obtain the symplectic product between two scalar-type modes that tend to zero
as $u\to\pm\infty$.

Similarly we find
\bea
2\overline{F^{(l)}_{rr}}\partial_u F^{(l')}_{uu}&-& 4\overline{F^{(l)}_{rr}}F^{(l')}_{uu} \approx 2\partial_u \Phi_S^{(l')}(\partial^2 _u +3\partial_u+2) \nonumber \\
&&\times \left(\partial^2 _r +n\partial_r+\frac{n(n-2)}{4}\right)\overline{\Phi_S^{(l)}},
\eea
so that
\begin{eqnarray}
&&2\overline{F^{(l)}_{rr}}\partial_u F^{(l')}_{uu}- 4\overline{F^{(l)}_{rr}}F^{(l')}_{uu}+ 2n(n-2)\overline{F^{(l)}}\partial_u F^{(l')}\nonumber \\
& & \approx 2\partial_u\Phi_S^{(l')}(\partial_u+1)(\partial_u +2)(\partial^2 _r +n\partial_r)\overline{\Phi_S^{(l)}}\nonumber\\
&&\ \ \ \ +\frac{3n(n-2)}{2}\partial_u\Phi_S^{(l')}\partial_u\overline{\Phi_{S}^{(l)}}\nonumber \\
&&\ \ \ \ +n(n-2)\overline{\Phi_S^{(l)}} \partial_u \Phi_S^{(l')} \nonumber\\
&&\ \ \ \  +\frac{n(n-2)}{2}\left(1-\frac{2}{n}-\frac{A_{n,l}}{n}\right)^2\overline{\Phi_S^{(l)}}\partial_u\Phi_S^{(l')}. \label{P1}
\end{eqnarray}

We can rewrite the first term, using Eq.~(\ref{horizon-field-eq}), as
\begin{eqnarray}
&&2\partial_u\Phi_S^{(l')}(\partial_u+1)(\partial_u +2)(\partial^2 _r +n\partial_r)\overline{\Phi_S^{(l)}} \nonumber \\
&& = 2\partial_u\Phi_S^{(l')}(\partial_u+1)(\partial_u +2)(\partial^2 _r +2\partial_r)\overline{\Phi_S^{(l)}} \nonumber\\
&&\ \ \ \  -(n-2)A_{n,l}\partial_u \Phi_S^{(l')}(\partial_u+2)\overline{\Phi_S^{(l)}}.
\end{eqnarray}
Substituting this equation into  Eq. (\ref{P1}), we find
\begin{eqnarray}
&& 2\overline{F^{(l)}_{rr}}\partial_u F^{(l')}_{uu}- 4\overline{F^{(l)}_{rr}}F^{(l')}_{uu}+ 2n(n-2)\overline{F^{(l)}}\partial_u F^{(l')} \nonumber \\
&&\approx 2\partial_u\Phi_S^{(l')}(\partial_u+1)(\partial_u +2)(\partial^2 _r +2\partial_r)\overline{\Phi_S^{(l)}} \nonumber\\
&&\ \ \ \ +\left[\frac{3n(n-2)}{2}-(n-2)A_{n,l}\right]\partial_u\Phi_S^{(l')}\partial_u\overline{\Phi_S^{(l)}} \nonumber\\
&&\ \ \ \  +\left[2-\frac{4A_{n,l}}{n}+\left(1-\frac{2}{n}-\frac{A_{n,l}}{n}\right)^2\right] \nonumber\\
&&\ \ \ \ \times \frac{n(n-2)}{2}\overline{\Phi_S^{(l)}} \partial_u \Phi^{(l')}.  \label{almost-there}
\end{eqnarray}

Now we note that
\begin{eqnarray}
\frac{1}{2}\left[\Box(r^2\Box\Phi_S^{(l)})-2\Box\Phi_S^{(l)}\right]&=&2(\partial_u+1)(\partial_u+2) \nonumber\\
&&\times(\partial^2 _r +2 \partial_r)\Phi_S^{(l)}. \label{box3}
\end{eqnarray}
To calculate $\Box(r^2\Box \Phi_S^{(l)})$, we write
\begin{eqnarray}
\Box \Phi_S^{(l)}=\frac{B_{n,l}+C_{n,l}r^2}{r^2}\Phi_S^{(l)}, \label{not-yet}
\end{eqnarray}
with
\beq
B_{n,l}=\frac{1}{4}\left[4 l(l+n-1)+n(n-2)\right]
\eeq
and
\beq
C_{n,l}=-\frac{(n-2)(n-4)}{4}.
\eeq
It is important not to let $r=1$ in Eq.~(\ref{not-yet}) because we are going to differentiate this expression
with respect to $r$. Then, we have, noting that $A_{n,l}=B_{n,l}+C_{n,l}$ for $r=1$,
\be
\Box(r^2\Box \Phi_S^{(l)})=(A_{n,l}^2-4C_{n,l})\Phi_S^{(l)}-4C_{n,l}\partial_u \Phi_S^{(l)}. \label{boxbox}
\ee
Substituting Eq.~(\ref{boxbox}) into Eq.~(\ref{box3}) and using the resulting expression in
Eq.~(\ref{almost-there}), we obtain
\begin{eqnarray}
&& 2\overline{F^{(l)}_{rr}}\partial_u F^{(l')}_{uu}- 4\overline{F^{(l)}_{rr}}F^{(l')}_{uu}
+ 2n(n-2)\overline{F^{(l)}}\partial_u F^{(l')} \nonumber \\
&&\approx \left\{\frac{A_{n,l}^2}{2}-A_{n,l}-2C_{n,l}+ \frac{n(n-2)}{2} \right. \nonumber\\
&&\ \ \ \ \times \left.\left[2-\frac{4A_{n,l}}{n}+\left(1-\frac{2}{n}-\frac{A_{n,l}}{n}\right)^2\right]\right\}\overline{\Phi_S^{(l)}}\partial_u \Phi_S^{(l')} \nonumber \\
&&\ \ \ \  +\left[\frac{3n(n-2)}{2}-(n-2)A_{n,l}-2C_{n,l}\right] \nonumber \\
&&\ \ \ \ \times \partial_u \Phi_S^{(l')} \partial_u \overline{\Phi_S^{(l)}}.
\end{eqnarray}
%With the aid of MatLab (http://www.mathworks.com/products/matlab/), we can simplify this expression, using symbolic %computation.
Substituting this equation into Eq. (\ref{scalar-inner}), we find for the inner product between
two scalar-type modes
\bea
\langle h^{(S;\omega l \sigma)}&,&h^{(S;\omega' l' \sigma')}\rangle=i\frac{(n-1)l(l-1)(l+n-1)(l+n)}{n} \nonumber\\
&&\times \int d\Omega_n du \overline{\mathbb{S}^{(l\sigma)}}\mathbb{S}^{(l'\sigma')} \nonumber\\
&&\times(\overline{\Phi_S^{(\omega l)}}\partial_u \Phi_S^{(\omega' l)} -\Phi_S^{(\omega' l)}\partial_u \overline{\Phi_S^{(\omega l)}}).
\eea
In $tr$ coordinates and on the $t=$\,constant Cauchy surface, this is given as
\bea
&&\langle h^{(S;\omega l \sigma)},h^{(S;\omega' l' \sigma')}\rangle=i\frac{(n-1)l(l-1)(l+n-1)(l+n)}{n} \nonumber\\
&&\times \delta^{l l'}\delta^{\sigma \sigma'} \int_{0}^{1} \frac{dr}{1-r^2} (\overline{\Phi_S^{(\omega l)}}\partial_t \Phi_S^{(\omega' l)} -\Phi_S^{(\omega' l)}\partial_t \overline{\Phi_S^{(\omega l)}}).
\eea

%%%%%%%%%%%%%%%%%%%%%%%%%%%%%%%%%%%%%%%%%%%%%%
\section{The Two-Point Function for the Scalar Field} \label{appendix:scalar-2p-function}
%%%%%%%%%%%%%%%%%%%%%%%%%%%%%%%%%%%%%%%%%%%%%%
The minimally-coupled scalar field equation with mass $M$,
\be
r^{-n} D_a (r^n D^a \phi) + \frac{1}{r^2}\hat{D}_i \hat{D}^i \phi - M^2\phi = 0,
\ee
can readily be solved with the positive-frequency solutions
being given by
\be
\phi^{(\omega l\sigma)}(y) =  N^{(\omega l)} e^{-i\omega t}r^l (1-r^2)^{i\omega/2}F(\alpha_-,\alpha_+;\gamma;r^2)
\mathbb{S}^{(l\sigma)},
\ee
where
\bea
\alpha_{\pm} & = & \frac{1}{2}\left( i\omega + l + \frac{n+1}{2} \pm \sqrt{\left( \frac{n+1}{2}\right)^2 - M^2}\right),
\nonumber \\ \\
\gamma& = & l +\frac{n+1}{2}.
\eea
The normalization constants $N^{(\omega l)}$  are determined by requiring
\bea
\langle \phi^{(\omega l \sigma)},\phi^{(\omega' l'\sigma')}\rangle
& : = & i \int_{\Sigma} d\Sigma n^\lambda
\overline{\phi^{(\omega l \sigma)}}\stackrel{\leftrightarrow}{\nabla}_\lambda \phi^{(\omega' l'\sigma')}
 \nonumber \\
& = & \delta^{ll'}\delta^{\sigma\sigma'}\delta(\omega - \omega').
\eea
Proceeding in exactly the same way as in the graviton case, we find
\be
|N^{(\omega l)}|^2 =
\dfrac{\sinh \pi\omega\left|\Gamma(\alpha_{-})\Gamma(\alpha_{+})\right|^2}
{4\pi^2 \left|\Gamma(l+\tfrac{n+1}{2})\right|^2}. \label{normal}
\ee
The special case with $n=2$ agrees with Ref.~\cite{higuchi}.

For $M>0$ we find that the normalization constants
$|N^{(\omega l)}|^2$ tend to $0$ like $\omega^{1/2}$ as $\omega \to 0$.  Now, for $M=0$
we have
\bea
&& \left. |N^{(\omega l)}|^2\right|_{M=0} \nonumber \\
& & = \dfrac{\sinh\pi\omega\left| \Gamma(\tfrac{1}{2}(i\omega + l))\Gamma(\tfrac{1}{2}(i\omega + l + n+1))\right|}
{ 4\pi^2\left|\Gamma(l+\tfrac{n+1}{2})\right|^2}.
\eea
Thus, the mode functions $\phi^{(\omega,l\sigma)}(y)$ tend to zero like $\omega^{1/2}$ for $l \geq 1$, but
the $l=0$ mode function diverges like $\omega^{-1/2}$.  The two-point function for the corresponding quantum field $\hat{\phi}(y)$ is
\bea
\langle \hat{\phi}(y)\hat{\phi}(y')\rangle
& = & \sum_{l=0}^\infty\sum_{\sigma}
\int_0^\infty d\omega \nonumber \\
&& \times \left[\frac{1}{e^{2\pi \omega} - 1}\overline{\phi^{(\omega l \sigma)}}(y)\phi^{(\omega l \sigma)}(y')\right.  \nonumber \\
&&\left. +\frac{1}{1 - e^{-2\pi\omega}}\phi^{(\omega l\sigma)}(y)\overline{\phi^{(\omega l\sigma)}}(y')\right].
\eea
This is IR divergent for $M=0$ because the $l=0$ contribution to the integrand behaves like $\omega^{-2}$ as
$\omega \to 0$.

%%%%%%%%%%%%%%%%%%%%%%%%%%%%%%%%%%%%%%%%%%%%%%
\section{Two-point function with one point at $r=0$}
\label{appendix}
%%%%%%%%%%%%%%%%%%%%%%%%%%%%%%%%%%%%%%%%%%%%%%
In this Appendix we show that the two-point function (\ref{two-point-result}) vanishes if one of the two points is at $r=0$.
This shows clearly that the values of the graviton two-point function by themselves have no physical significance.

Since $r=0$ is a coordinate singularity of spherical polar coordinates, we need to contract the indices of the two-point
function at the origin with \textit{vielbein} ${e_{(a)}}^\mu$ satisfying
\be
{e_{(a)}}^{\mu} {e_{(b)}}^{\nu} \eta^{(a)(b)} = g^{\mu\nu}
\ee
and
\be
 g_{\mu\nu}{e_{(a)}}^{\mu}{e_{(b)}}^{\nu} = \eta_{(a)(b)},
\ee
where $\eta_{(a)(b)}=\textrm{diag}(-1,1,1,...,1)$.
At any point away from $r=0$ we can choose the following vielbein $e_{(a)}^\mu$:
\begin{eqnarray}
\hat{e}_{(0)} &=& \left((1-r^2)^{-1/2},0,...,0 \right), \\
\hat{e}_{(1)} &=& \left(0,(1-r^2)^{1/2},0,...,0\right), \\
\hat{e}_{(i)} &=& \left(0,0,...,\frac{1}{r\sqrt{\gamma_{ii}}},0,...,0\right),
\end{eqnarray}
where the index $i$ is not summed over.  We take the limit $r\to 0$ after contracting the indices of the two-point function
at the origin with this vielbein.

Now we examine the components ${e_{(a)}}^{\mu}{e_{(b)}}^{\nu}h^{(P;\omega l \sigma)}_{\mu\nu}(y)$ as $r$ in
$y=(t,r,\theta,\phi,...)$ tends to zero. If $(a)$ and $(b)$ are $(0)$ or $(1)$, then
$\lim_{r\to 0}\hat{e}_{(a)}^\mu \hat{e}_{(b)}^\nu h_{\mu\nu} = h_{ab}$, where $a$ and $b$ on the right-hand side are
$t$ or $r$.  Hence we can examine the components $h_{ab}$ directly.
For the vector- and tensor-type perturbations this is trivially zero since $h^{(P;\omega l \sigma)}_{ab}=0$ for $P=V$ and $T$ in the gauge that we have chosen.  For the scalar-type modes, we first note that $r^{n/2}\Psi_S^{(\omega l)}$ in Eq.~(\ref{S_ab})
behaves like $r^{l+n}$ as $r\to 0$.  The derivative operators $D_aD_b$ and $\Box$ change the leading behavior to
 $O(r^{l+n-2})$.  Then it can readily be seen that $h_{ab}^{(S;\omega l\sigma)}$ tends to zero like $r^{l}$ ($l\geq 2$) or faster as $r\to 0$.

For $(a)=(0)$ or $(1)$ and $(b)=(i)$, we find
\be
{e_{(a)}}^{\mu}{e_{(i)}}^{\nu}h^{(P;\omega l \sigma)}_{\mu\nu}=(1-r^2)^{\textrm{sign}(a)/2}(r^2\gamma_{ii})^{-1/2}h^{(P;\omega l \sigma)}_{a i}, \label{vielbeineqV}
\ee
where $\textrm{sign}(a)=-1$ if $a=0$ and $\textrm{sign}(a)=1$ if $a=1$.
Now, it is the tensor- and scalar-type perturbations that vanish identically in the gauge we have chosen.
For the vector case,  we find that $r^{n/2}\Phi_V^{(\omega l)}$ in Eq.~(\ref{V_ai}) behaves like $r^{l+n}$
as $r\to 0$. Then it can readily be seen that $h_{ti}^{(V;\omega l \sigma)}$ and
$h_{ri}^{(V;\omega l \sigma)}$ behave like $r^{l+1}$ and $r^{l+2}$, respectively, with $l\geq 2$.  Then
Eq.~(\ref{vielbeineqV}) shows that ${e_{(a)}}^{\mu}{e_{(i)}}^\nu h^{(P;\omega l\sigma)}_{\mu\nu}\to 0$
as $r\to 0$.
%\be
%{e_{(0)}}^{\mu}{e_{(i)}}^{\nu}h^{(\omega l \sigma)}_{\mu\nu}=\sqrt{\frac{1-r^2}
%{\gamma_{ii}}}\frac{\mathbb{V}^{(l \sigma)}_i}{r^{n-1}}\frac{\partial}{\partial r}(r^{n/2}\Phi_V^{(\omega l
%\sigma)}).
%\ee
%This will be proportional to $r^{l}$ again, so it will vanish when $r \rightarrow 0$.
%
%For the \textit{vielbein} component $(a)=(1), (b)=(i)$,
%\be
%{e_{(1)}}^{\mu}{e_{(i)}}^{\nu}h^{(\omega l \sigma)}_{\mu\nu}=(1-r^2)^{1/2}
%(r^2\gamma_{ii})^{-1/2}h^{(\omega l
%\sigma)}_{ri}.
%\ee
%The vector-type perturbation will be
%\be
%{e_{(1)}}^{\mu}{e_{(i)}}^{\nu}h_{\mu\nu}=-i\omega\frac{\mathbb{V}_i}{(1-r^2)^{1/2}
%(\gamma_{ii})^{1/2}}\frac{\Phi_V}{r^{\frac{n}{2}-1}} \propto r^{l+1}.
%\ee
%So it will vanish for $r \rightarrow 0$. The tensor and scalar cases will vanish identically.

Finally, we calculate the components with $(a)=(i)$ and $(b)=(j)$ to find
\be
{e_{(i)}}^{\mu}{e_{(j)}}^{\nu}h^{(P;\omega l \sigma)}_{\mu\nu}=(\gamma_{ii}\gamma_{jj})^{-1/2}r^{-2}h^{(P;\omega l \sigma)}_{ij}.
\ee
The vector case is trivial since $h_{ij}^{(V;\omega l\sigma)}=0$. The scalar case is
\beq
\hspace{-1cm}{e_{(i)}}^{\mu}{e_{(j)}}^{\nu}h^{(S;\omega l \sigma)}_{\mu\nu}=\frac{\gamma_{ij}\mathbb{S}^{(l \sigma)}}{(\gamma_{ii}\gamma_{jj})^{1/2}}\frac{(\Box +2)}{nr^{n-2}}(r^{n/2}\Phi_S^{(\omega l)}),
\eeq
which behave like $r^{l}$, $l\geq 2$, as $r\to 0$.
For the tensor case, we have
\beq
{e_{(i)}}^{\mu}{e_{(j)}}^{\nu}h^{(T;\omega l \sigma)}_{\mu\nu}=\frac{2\mathbb{T}^{(l \sigma)}_{ij}}{(\gamma_{ii}\gamma_{jj})^{1/2}}r^{-n/2}\Phi^{(\omega l)}_T.
\eeq
Then these vielbein components for the tensor case behave like $r^{l}$, $l\geq 2$, for small $r$.
Hence, it will vanish as $r \rightarrow 0$.
Since all these components vanish for $r \to 0$, the two-point function itself vanishes in this limit.

%%%%%%%%%%%%%%%%%%%%%%%%%%%%%%%%%%%%%%%%%%%
% Bibliography
%\bibliographystyle{h-physrev4}
%\bibliographystyle{myutphys}
%\bibliographystyle{abbrv}
\bibliography{refs_HBC}
\end{document}